\documentclass[aps,prb,reprint,showpacs,amsmath,amssymb]{revtex4-1}
\usepackage{graphicx}
\usepackage{bm}

\begin{document}

\title{Extrinsic spin Nernst effect in two-dimensional electron systems}

\author{Hiroshi \textsc{Akera} and Hidekatsu \textsc{Suzuura}}

\affiliation{Division of Applied Physics, Faculty of Engineering, 
Hokkaido University, Sapporo, Hokkaido, 060-8628, Japan}

\date{\today}

\begin{abstract}
The spin accumulation due to the spin current induced by 
the perpendicular temperature gradient (the spin Nernst effect)  
is studied 
in a two-dimensional electron system (2DES) with spin-orbit interaction 
by employing the Boltzmann equation. 
The considered 2DES is confined within a symmetric quantum well 
with delta doping at the center of the well.   
A symmetry consideration leads to 
the spin-orbit interaction which  
is diagonal in the spin component perpendicular to the 2DES. 
As origins of the spin current, 
the skew scattering and the side jump are considered 
at each impurity on the center plane of the well. 
It is shown that, for repulsive impurity potentials, 
the spin-Nernst coefficient changes its sign 
at the impurity density where 
contributions from the skew scattering and the side jump 
cancel each other out.  
This is in contrast to the spin Hall effect 
in which the sign change of the coefficient occurs  
for attractive impurity potentials.
\end{abstract}

\pacs{72.25.Dc, 73.63.Hs}


\maketitle

\def\sxx{\sigma_{xx}}
\def\sxy{\sigma_{xy}}
\def\syx{\sigma_{yx}}
\def\syy{\sigma_{yy}}
\def\oc{\omega_{\rm c}}
\def\ve{\varepsilon}
\def\Te{T_{\rm e}}
\def\muec{\mu_{\rm ec}}
\def\kB{k_{\rm B}}
\def\Vec#1{\bm{#1}}

\section{Introduction}

The spin Hall effect\cite{Sinova_et_al_2006, Dyakonov2008, Vignale2010} is 
the generation of the spin accumulation, or 
the difference in density between spin-up and spin-down electrons,  
due to the spin current driven by the perpendicular electric field.  
This transverse effect is produced by spin-orbit interaction 
in the absence of magnetic field.  
It has attracted much attention in the field of spintronics\cite{Zutic2004spintronics} 
as a promising way to create the spin accumulation in nonmagnetic materials. 
The first report on the observation of the spin Hall effect 
has been made by Kato et al.\cite{Kato2004} 
for three-dimensional electron systems (3DES) in semiconductors, 
n-doped GaAs and n-doped InGaAs, 
and is followed by 
many experimental works including the observation 
in two-dimensional hole systems (2DHS)\cite{Wunderlich2005} 
and that in two-dimensional electron systems (2DES).\cite{Sih2005} 
Theoretical proposals have been made before such observations 
and are classified into the intrinsic origin and the extrinsic one.   
The intrinsic spin Hall effect\cite{Murakami2003, Sinova2004} 
is due to the spin-orbit interaction induced by the crystal potential 
as well as the confining potential of a quantum well.   
The extrinsic spin Hall effect\cite{Dyakonov1971JETPL, Dyakonov1971PhysicsLettersA, Hirsch1999, Zhang2000} 
originates from electron scatterings from nonmagnetic impurities 
in the presence of the spin-orbit interaction. 
The spin Hall effect observed in the 3DES\cite{Kato2004} and that in the 2DES\cite{Sih2005} 
have been explained by calculations based on the extrinsic mechanism.\cite{Engel2005, Hankiewicz2006PRB, Tse-Das_Sarma2006}
In this paper we investigate the extrinsic spin Nernst effect in 2DES. 

The observation of the spin Hall effect in 2DES has been made by Sih et al.\cite{Sih2005} 
in a (110) AlGaAs quantum well. 
They have already suggested in their paper that the observed spin Hall effect is extrinsic, 
since (1) the quantum well is doped at the area density of $10^{12}$cm$^{-2}$,  
(2) the measured value of the Rashba coefficient is small, and 
(3) the Dresselhaus field should be absent because of 
the current orientation along the [001] axis in the (110) quantum well. 
Since the measurement in 2DES,\cite{Sih2005} as the 3DES experiment,\cite{Kato2004} 
is made at the temperature of 30K,  
the phase coherence in the electron transport may not be important. 
Therefore a theoretical study for this experiment has been performed 
based on the Boltzmann equation 
by Hankiewicz and Vignale,\cite{Hankiewicz2006PRB} 
as well as the semiclassical theory by Engel et al.\cite{Engel2005} for the 3DES experiment. 

In the atomic-layer epitaxial growth of a semiconductor heterostructure, 
both positively and negatively ionized impurities 
can be introduced at a precise distance from the heterointerface  
by employing the method of delta-doping.\cite{Ploog1987}
In fact, 
both Si (donor) and Be (acceptor) have been doped successfully
at a precise distance from the interface of a GaAs/AlGaAs heterostructure,  
and a strong dependence on the dopant type 
has been found in magnetotransport properties 
of a 2DES located near the dopant.\cite{Haug1987} 
Such an accurate control of the doping profile 
gives the 2DES an advantage 
in that this can provide a method to enhance strongly 
the spin accumulation due to the extrinsic spin Hall effect. 

A remarkable dependence of the extrinsic spin Hall current 
on the impurity-limited mobility has been found in a model of 2DES 
by Hankiewicz and others.\cite{Hankiewicz2006PRB, Hankiewicz2006PRL}  
The 2DES in their model has a negligible width and therefore 
the dependence on the above-mentioned doping profile is beyond the scope of their works. 
There are two contributions to the extrinsic spin Hall current. 
One is the contribution from the skew scattering\cite{Mott1929, Smit1955, Smit1958} 
and the other is that from the side jump.\cite{Berger1970, Berger1972, Lyo-Holstein1972} 
Both have long been studied in the theory of 
the anomalous Hall effect in ferromagnetic metals 
(see Refs.\onlinecite{Karplus-Luttinger1954, Luttinger1958, Nozieres1973} 
for early theories on the anomalous Hall effect 
and Ref.\onlinecite{Nagaosa2010} for a recent review). 
The skew-scattering contribution has a different sign 
depending on whether the impurity potential is attractive or repulsive, 
while the side-jump contribution is independent 
of both the impurity potential and the impurity density. 
For attractive impurity potentials, 
the contributions from the skew scattering and the side jump are opposite in sign.  
Therefore the direction of the spin current is switched  
as the weight of the skew-scattering contribution is changed, for example 
by varying the mobility.\cite{Hankiewicz2006PRB, Hankiewicz2006PRL}  
This theoretical finding suggests that 
the spin accumulation due to the extrinsic spin Hall effect 
can be controlled in a wide range, for example, by changing the impurity density.  
We expect that the controllability should be enhanced 
by introducing various doping profiles with the delta-doping technique.

The temperature gradient is another driving force for 
the spin current in the perpendicular direction. 
This phenomenon, called the spin Nernst effect, 
is one of the most important subjects in `spin caloritronics', 
a research field exploring the interplay between the heat and the spin degree of freedom,\cite{Johnson1987,Bauer2012} 
The spin Nernst effect is the nonmagnetic analogue of the anomalous Nernst effect. 
While the anomalous Nernst effect has been studied in 3D ferromagnetic metals for nearly a century 
(see Refs.\onlinecite{Smith1911,Kondorskii_Vasileva1964} for early experiments,  
Refs.\onlinecite{Kondorskii1964,Berger1972} for early theories),   
studies on the spin Nernst effect have started quite recently. 
An experimental study to observe the spin Nernst effect 
is in progress in 3D metals.\cite{Seki2010} 
Several theoretical studies on the spin Nernst effect have been made in 2DES.\cite{Cheng2008,Ma2010,Liu2010} 
However, these theories are only for the intrinsic origin due to the Rashba term. 
The spin Nernst effect with the extrinsic origin is worth studying theoretically, 
in particular, the dependence on the type and the density of impurities.  
Even the sign of each contribution in the extrinsic mechanism is 
not known in the spin Nernst effect. 

In this paper we study theoretically 
the spin Nernst effect in 2DES based on the extrinsic mechanism  
by employing the Boltzmann equation.  
In particular, we propose an efficient method to control the spin Nernst effect 
by changing the impurity type and density. 

In Sec.~\ref{sec: Formulation} we describe our formulation.
We start from the Hamiltonian for an electron in a quantum well 
formed in a semiconductor heterostructure 
with interfaces parallel to the $xy$ plane. 
Then we reduce it to the effective Hamiltonian for the two-dimensional electron motion 
in the $xy$ plane (Sec.~\ref{sec: 2D_Hamiltonian}). 
Here we show that the 2D Hamiltonian becomes diagonal in the $z$ component of spin 
when each impurity is located on the center plane of a symmetric quantum well. 
For such 2D Hamiltonian we write the Boltzmann equation 
and derive the distribution function (Sec.~\ref{sec: Boltzmann_equation}). 
Using the distribution function we obtain 
the current densities and the transport coefficients (Sec.~\ref{eq:current_densities}). 
We show here that the side jump also gives rise to the current density component 
induced by the temperature gradient. 

Then we apply the formulation to the spin Nernst effect in Sec.~\ref{sec:spin_Nernst}.
We consider a rectangular 2DES, 
apply the temperature gradient along the $x$ direction,  
and calculate the gradient along $y$ of 
the chemical-potential difference between spin-up and spin-down electrons.  
We pay a special attention to the signs of contributions 
from the skew scattering and the side jump. 
We present the result as a function of the impurity density 
for both attractive and repulsive potentials
and compare it with that of the spin Hall effect. 
Conclusions are given in Sec.~\ref{eq:conclusions}.

\section{Formulation}
\label{sec: Formulation}
\subsection{2D Hamiltonian}
\label{sec: 2D_Hamiltonian}

We consider conduction-band electron states  
which are bound to a quantum well with translational symmetry in the $xy$ plane.  
We assume that the wave function describing the motion along the $z$ direction 
is frozen to the ground state,  
and derive the effective Hamiltonian for the 2D motion in the $xy$ plane 
in the following.  

We start from the Hamiltonian describing the 3D motion:  
\begin{equation}
 H_{\rm 3D} = 
\frac{ p_x^2 +  p_y^2 +  p_z^2}{2m} 
+ V_{\rm 3D}(x,y,z) 
- \alpha {\Vec \sigma} \cdot 
\left(\Vec \nabla V_{\rm 3D} \times {\Vec p} \right) 
 ,
\label{eq:Hamiltonian_3D}
\end{equation}
where $m$ is the effective mass,  
$\alpha$ is the effective coupling constant of the spin-orbit interaction 
for an electron in the conduction band of the semiconductor, 
and ${\Vec \sigma}=(\sigma_x,  \sigma_y, \sigma_z)$ is 
the Pauli spin matrix. 
The potential energy is 
\begin{equation}
V_{\rm 3D}(x,y,z) 
= V_{\rm well}(z) 
+ V_{\rm imp}(x,y,z)
+ e \Vec E \cdot \Vec r 
,
\label{eq:potential_3D}
\end{equation}
where $V_{\rm well}(z)$ is the well potential, 
$V_{\rm imp}(x,y,z)$ is the potential due to randomly-distributed impurities, 
$\Vec E =(E_x, E_y, 0)$ is the in-plane electric field, and
$e>0$ is the absolute value of the electronic charge.  

We define the Hamiltonian for two-dimensional motion as 
\begin{equation}
 H_{\rm 2D} = \left<  H_{\rm 3D} \right>
,
\label{eq:Hamiltonian_2D_from_3D}
\end{equation}
where the brackets represent the average with respect to the motion along $z$ as   
\begin{equation}
\left<  H_{\rm 3D} \right> = \int dz \ \varphi_0 (z)  H_{\rm 3D} \varphi_0 (z)
. 
\label{eq:average_z}
\end{equation}
Here $\varphi_0 (z)$ is the wave function of the ground state at energy $\ve_0$ which satisfies 
the Schr{\"o}dinger equation: 
\begin{equation}
\left[ \frac{ p_z^2}{2m} + V_{\rm well}(z) \right] \varphi_0 (z)
= \ve_0 \varphi_0 (z)
. 
\label{eq:ground_state}
\end{equation}

We begin with evaluating terms in $ H_{\rm 2D}$ 
which originate from the spin-orbit interaction. 
We here assume that $V_{\rm well}(z)$ is symmetric with respect to 
the center of the well, $z=0$. 
Then $V_{\rm well}(z)$ gives no spin-orbit term in $ H_{\rm 2D}$.
The in-plane electric field gives a spin-orbit term with $ \sigma_z$ only 
(no terms with $ \sigma_x$ and $ \sigma_y$), 
since $\left<  p_z \right> =0$ and $E_z=0$. 

Spin-orbit terms in $ H_{\rm 2D}$, which is due to the impurity potential, 
are separated into the following three components: 
\begin{equation}
\begin{split}
 H_{{\rm 2D},x}^{\rm so,imp} &= 
- \alpha  \sigma_x \left[\left<(\nabla_y V_{\rm imp})  p_z\right> 
- \left<\nabla_z V_{\rm imp}\right>  p_y \right] 
 ,\\
\label{eq:spin_orbit_impurity}
 H_{{\rm 2D},y}^{\rm so,imp} &= 
- \alpha  \sigma_y \left[\left<\nabla_z V_{\rm imp}\right>  p_x - \left<(\nabla_x V_{\rm imp})  p_z \right>\right] 
 ,\\
 H_{{\rm 2D},z}^{\rm so,imp} &=
- \alpha  \sigma_z \left[(\nabla_x v_{\rm imp})  p_y - (\nabla_y v_{\rm imp})  p_x \right] 
 ,
\end{split}
\end{equation}
with the effective impurity potential in the 2DES, 
\begin{equation}
v_{\rm imp}(x,y) = \left< V_{\rm imp}(x,y,z) \right>
.
\label{eq:potential_2D_from_3D}
\end{equation}
Since a term in $ H_{{\rm 2D},x}^{\rm so,imp}$ can be rewritten as   
$\left<(\nabla_y V_{\rm imp})  p_z\right> = i \hbar (\nabla_y \left<\nabla_z V_{\rm imp}\right>)/2$ 
and the same is true for $ H_{{\rm 2D},y}^{\rm so,imp}$, 
the magnitude of $ H_{{\rm 2D},x}^{\rm so,imp}$ and that of $ H_{{\rm 2D},y}^{\rm so,imp}$
are determined by 
$\left<\nabla_z V_{\rm imp}\right>$, 
$\nabla_x \left<\nabla_z V_{\rm imp}\right>$ and $\nabla_y \left<\nabla_z V_{\rm imp}\right>$.  
On the other hand the magnitude of $ H_{{\rm 2D},z}^{\rm so,imp}$ is determined by $\nabla_x v_{\rm imp}$ and $\nabla_y v_{\rm imp}$. 

Equation (\ref{eq:spin_orbit_impurity}) demonstrates that 
the 2D Hamiltonian for 2DES formed in a quantum well, in general, 
contains in-plane components of spin, $ \sigma_x$ and $ \sigma_y$, 
due to the combined action of the impurity potential and the spin-orbit interaction.  
The resulting spin relaxation due to the Elliott-Yafet mechanism\cite{Elliott1954, Yafet1963, Zutic2004spintronics}
has already been reported in the literature.\cite{Averkiev2002, Bronold2004} 
However, 
the $z$ component of spin, $ \sigma_z$, is conserved when the condition
\begin{equation}
\left<\nabla_z V_{\rm imp}\right>= 0
,
\label{eq:condition_for_2D}
\end{equation}
is satisfied. 
This condition is satisfied when impurities are located 
on the center plane ($z=0$) of the symmetric quantum well. 
Such a precise placement of impurities is 
in fact possible by using the method  of delta-doping.\cite{Ploog1987,Haug1987}

We therefore assume the condition Eq.(\ref{eq:condition_for_2D}). 
Our Hamiltonian for the two-dimensional motion of the 2DES 
is simplified to become  
\begin{equation}
 H_{\rm 2D} = 
\frac{ p_x^2 +  p_y^2}{2m} 
+ v_{\rm 2D}(x,y)
- \alpha  \sigma_z 
[(\nabla_x v_{\rm 2D})  p_y - (\nabla_y v_{\rm 2D})  p_x ] 
 ,
\label{eq:Hamiltonian_2D}
\end{equation}
with 
\begin{equation}
v_{\rm 2D}(x,y) 
= v_{\rm imp}(x,y)
+ e (E_x x + E_y y) 
.
\label{eq:potential_2D}
\end{equation}
This 2D Hamiltonian coincides with that 
employed to study the extrinsic spin Hall effect of 2DES 
in the previous theory.\cite{Hankiewicz2006PRB}

\subsection{Boltzmann equation and the distribution function}
\label{sec: Boltzmann_equation}

Hankiewicz and Vignale
in their study on the extrinsic spin Hall effect of 2DES\cite{Hankiewicz2006PRB} have 
obtained the distribution function by solving the Boltzmann equation 
up to the first order of the electric field $\Vec E$ and of the spin-orbit coupling constant $\alpha$. 
Here we extend their formulation to include gradients 
of the chemical potential and the electron temperature as driving forces,  
and obtain the distribution function 
up to the first order of all the driving forces, which is denoted simply by $O(E)$ below, 
and up to $O(\alpha)$.
We show that the side jump, as well as the skew scattering, gives a temperature-gradient term in the distribution function. 

Since our 2D Hamiltonian conserves 
the $z$ component of spin, 
the distribution function for each of its eigenvalues $\sigma=\pm 1$ 
is determined independently by the Boltzmann equation. 
The Boltzmann equation for the distribution function 
of electrons with spin $\sigma$, 
$f_{\sigma}(\Vec r, \Vec k)$ in a steady state is 
\begin{equation}
\Vec v \cdot \frac{\partial f_{\sigma}}{\partial \Vec r}
+ \frac{(-e)\Vec E}{\hbar} \cdot \frac{\partial f_{\sigma}}{\partial \Vec k}
=
\left(
\frac{\partial f_{\sigma}}{\partial t}
\right)_{\rm c}
.
\label{eq:Boltzmann_eq}
\end{equation}
The distribution function is decomposed into 
that in the local equilibrium, $f^{(0)}$, 
which depends on $\Vec k$ through the energy $\ve_k = \hbar^2 k^2 /2m$,  
and the deviation in the first order of the driving forces, $f_{\sigma}^{(1)}$, 
which depends on the direction of $\Vec k$ relative to $\Vec E$: 
\begin{equation}
f_{\sigma}(\Vec r, \Vec k) 
= f^{(0)}\left(\ve_k, \mu_{\sigma}(\Vec r), \Te(\Vec r)\right)
+ f_{\sigma}^{(1)}(\Vec r, \Vec k) 
,
\label{eq:f_sigma_k}
\end{equation}
where 
$f^{(0)}(\ve, \mu, T) = \{\exp[(\ve - \mu)/\kB T ]+1\}^{-1}$, 
$\mu_{\sigma}$ is the spin-dependent chemical potential, 
and $\Te$ is the electron temperature. 
Note that the first term of $f_{\sigma}(\Vec r, \Vec k)$ 
includes spatial dependences of $\mu_{\sigma}$ and $\Te$,  
although the function $f^{(0)}$ itself is 
of the zeroth order of the driving forces. 
The $\Vec r$ dependence of $f_{\sigma}^{(1)}(\Vec r, \Vec k)$ also 
originates from the driving forces, and therefore 
it gives only terms of $O(E^2)$. 
Since
\begin{equation}
\Vec v = \frac{\hbar \Vec k}{m} + O(E)
, \ \ \ 
\frac{1}{\hbar}
\frac{\partial f_{\sigma}}{\partial \Vec k} 
= \Vec v \ \frac{\partial f^{(0)}}{\partial \ve_k} + O(E)
,
\end{equation}
and
\begin{equation}
\frac{\partial f_{\sigma}}{\partial \Vec r} 
= 
\frac{\partial f^{(0)}}{\partial \mu_{\sigma}} 
\Vec \nabla \mu_{\sigma}
+
\frac{\partial f^{(0)}}{\partial \Te} 
\Vec \nabla \Te
 + O(E^2)
,
\end{equation}
then the left hand side of the Boltzmann equation 
Eq.(\ref{eq:Boltzmann_eq}) is written in the first order of the driving forces as 
\begin{equation}
\Vec v \cdot \frac{\partial f_{\sigma}}{\partial \Vec r}
+ \frac{(-e)\Vec E}{\hbar} \cdot \frac{\partial f_{\sigma}}{\partial \Vec k}
=
\Vec v \cdot \Vec F_{\sigma}(\ve_k) 
\frac{\partial f^{(0)}}{\partial \ve_k}
,
\label{eq:Boltzmann_eq_LHS}
\end{equation}
with a generalized force
\begin{equation}
\Vec F_{\sigma}(\ve_k)
= 
- \Vec \nabla \mu_{\sigma}^{\rm ec}
- \frac{\ve_k - \mu_{\sigma}}{\Te} \ \Vec \nabla \Te
. 
\label{eq:generalized_force}
\end{equation}
Here $\mu_{\sigma}^{\rm ec}$ is the spin-dependent electrochemical potential defined by
\begin{equation}
\mu_{\sigma}^{\rm ec}= e \Vec E \cdot \Vec r + \mu_{\sigma},  
\end{equation}
and the chemical potential $\mu_{\sigma}$ consists of terms 
in the zeroth and first orders of the driving forces: 
\begin{equation}
\mu_{\sigma} = \mu_{\sigma}^{(0)} + \mu_{\sigma}^{(1)}.  
\end{equation}

The collision term is written as\cite{Hankiewicz2006PRB}  
\begin{equation}
\left(
\frac{\partial f_{\sigma}}{\partial t}
\right)_{\rm c}
=
 \sum_{\Vec k'} 
\left[
- W_{\Vec k \Vec k' \sigma} f_{\sigma}(\Vec k) 
+ W_{\Vec k' \Vec k \sigma} f_{\sigma}(\Vec k') 
\right] ,
\end{equation}
where $W_{\Vec k \Vec k' \sigma}$ is the rate of transition 
from $\Vec k \sigma$ to $\Vec k' \sigma$ 
and has the contribution from the normal scattering, $W^{\rm n}_{\Vec k \Vec k' \sigma}$, and 
that from the skew scattering, $W^{\rm ss}_{\Vec k \Vec k' \sigma}$: 
\begin{equation}
W_{\Vec k \Vec k' \sigma} 
= 
W^{\rm n}_{\Vec k \Vec k' \sigma} + W^{\rm ss}_{\Vec k \Vec k' \sigma} 
\end{equation}
with
\begin{equation}
\begin{split}
&W^{\rm n}_{\Vec k \Vec k' \sigma}
=
W_{\rm n}(\ve_k, \theta) 
\delta (\ve_{k'} - \ve_k + e \Vec E \cdot \Delta \Vec r), \\
& W^{\rm ss}_{\Vec k \Vec k' \sigma}
=
\sigma \sin \theta \ W_{\rm ss}(\ve_k, \theta)
\delta (\ve_{k'} - \ve_k + e \Vec E \cdot \Delta \Vec r),
\end{split}
\end{equation}
Here $\theta$ is the angle of $\Vec k'$ relative to that of $\Vec k$. 
Since we retain only terms up to $O(\alpha)$,  
$W_{\rm ss}(\ve_k, \theta)$ representing the skew scattering is $O(\alpha)$, 
while $W_{\rm n}(\ve_k, \theta)$ due to the normal scattering 
has no dependence on $\alpha$.  
Both $W_{\rm n}(\ve_k, \theta)$ and $W_{\rm ss}(\ve_k, \theta)$ are an even function of $\theta$. 
The delta function expresses the conservation of energy,  
in which we take into account 
the potential energy shift due to 
the position change in the side jump at the scattering 
from $\Vec k \sigma$ to $\Vec k' \sigma$, 
\begin{equation}
\Delta \Vec r = -2 \alpha \sigma \hbar (\Vec k' - \Vec k) \times \Vec e_z
, 
\end{equation}
where $\Vec e_z =(0,0,1)$ and 
the vector $\Vec k$ should be regarded as a three-dimensional vector 
with vanishing $z$ component, $\Vec k =(k_x,k_y,0)$. 
Note that the functions $W_{\rm n}(\ve_k, \theta)$ and $W_{\rm ss}(\ve_k, \theta)$
are defined in the absence of $\Vec E$ 
where the difference between $\ve_k$ and $\ve_{k'}$ is absent. 

The collision term is separated into four components, 
\begin{equation}
\left(
\frac{\partial f_{\sigma}}{\partial t}
\right)_{\rm c}
=
C_{\rm n0} + C_{\rm n1} + C_{\rm ss0} + C_{\rm ss1}
,
\end{equation}
with 
\begin{equation}
\begin{split}
&C_{\rm n0} = \sum_{\Vec k'} W^{\rm n}_{\Vec k \Vec k' \sigma}
 \left[ f^{(0)}(\ve_{k'}) - f^{(0)}(\ve_k) \right]
,\\
&C_{\rm n1} = \sum_{\Vec k'} W^{\rm n}_{\Vec k \Vec k' \sigma}
\left[ f_{\sigma}^{(1)}(\Vec k') - f_{\sigma}^{(1)}(\Vec k) \right]
,\\
&C_{\rm ss0} = \sum_{\Vec k'} W^{\rm ss}_{\Vec k \Vec k' \sigma} 
\left[ - f^{(0)}(\ve_{k'}) - f^{(0)}(\ve_k) \right]
,\\
&C_{\rm ss1} = \sum_{\Vec k'} W^{\rm ss}_{\Vec k \Vec k' \sigma} 
\left[ - f_{\sigma}^{(1)}(\Vec k') - f_{\sigma}^{(1)}(\Vec k) \right]
. 
\end{split}
\end{equation}
We retain terms up to $O(E)$ and those up to $O(\alpha)$.   
Then we immediately have $C_{\rm ss0}=0$ 
since the side jump $\Delta \Vec r $ giving terms of $O(\alpha^2)$ in $C_{\rm ss0}$ is to be neglected 
and the integrand of $C_{\rm ss0}$ becomes an odd function of $\theta$. 
On the other hand, $C_{\rm n0}$ is not zero in the presence of the side jump. 
The side jump gives 
the difference between $f^{(0)}(\ve_{k'})$ and $f^{(0)}(\ve_k)$ of $C_{\rm n0}$ 
in two ways. 
One is from the difference in the kinetic energy $\ve_k$,  
which comes from the potential energy shift and the energy conservation at the scattering. 
The other is from the difference in the distribution 
between two points separated by $\Delta \Vec r$, 
which is described in the local equilibrium by 
the difference in $\mu_{\sigma}$ and that in $\Te$.  
Such considerations give 
\begin{eqnarray}
f^{(0)}(\ve_{k'}) - f^{(0)}(\ve_k)
&=& \frac{\partial f^{(0)}}{\partial \ve_k} (\ve_{k'}-\ve_k) 
+ \frac{\partial f^{(0)}}{\partial \mu_{\sigma}} \Vec \nabla \mu_{\sigma} \cdot \Delta \Vec r
\nonumber \\ 
&+& \frac{\partial f^{(0)}}{\partial \Te} \Vec \nabla \Te \cdot \Delta \Vec r
, 
\end{eqnarray}
and $\ve_{k'}-\ve_k = -e \Vec E \cdot \Delta \Vec r$ 
using the energy conservation. 

We seek the solution for $f_{\sigma}^{(1)}$ of the form 
\begin{equation}
f_{\sigma}^{(1)}(\Vec k)
= - \frac{\partial f^{(0)}}{\partial \ve_k} 
\hbar \Vec k \cdot \Vec V_{\sigma} (\ve_k), 
\label{eq:f_1}
\end{equation}
and substitute this form into $C_{\rm n1}$ and $C_{\rm ss1}$.  
Then a straightforward calculation gives, for $\ve=\ve_k$,  
\begin{eqnarray}
\hskip -0.7cm\left(
\frac{\partial f_{\sigma}}{\partial t}
\right)_{\rm c}
&&= \frac{\partial f^{(0)}}{\partial \ve} \hbar \Vec k
\nonumber \\ 
&&
\cdot 
\left[
\frac{\Vec V_{\sigma}(\ve)}{\tau_{\rm n}(\ve)} 
+
\frac{\sigma \Vec V_{\sigma}(\ve) \times \Vec e_z}{\tau_{\rm ss}(\ve)} 
+
\frac{2\alpha \sigma \Vec e_z \times \Vec F_{\sigma}(\ve)}{\tau_{\rm n}(\ve)}
\right]
. 
\label{eq:Boltzmann_eq_RHS}
\end{eqnarray}
The first and second terms in the square brackets 
come from $C_{\rm n1}$ (the normal scattering) 
and $C_{\rm ss1}$ (the skew scattering), respectively, 
with $\tau_{\rm n}$ and $\tau_{\rm ss}$ 
defined by 
\begin{eqnarray}
\frac{1}{\tau_{\rm n}(\ve)}
&=& 
\sum_{\Vec k'} 
\delta (\ve_{k'} - \ve) W_{\rm n}(\ve, \theta) (1-\cos \theta) 
, \\
\frac{1}{\tau_{\rm ss}(\ve)}
&=& 
\sum_{\Vec k'}  
\delta (\ve_{k'} - \ve) W_{\rm ss}(\ve, \theta) \sin^2 \theta 
. 
\end{eqnarray}
Note that $\tau_{\rm ss}(\ve)$ can be negative 
since $W_{\rm ss}(\ve, \theta)$ starts from the third order 
in the expansion with respect to the impurity potential.\cite{Landau1965} 
The third term comes from $C_{\rm n0}$ (the side jump) 
and is induced by the gradient of the chemical potential and 
that of the electron temperature as well as the electric field. 
Substituting the drift term Eq.(\ref{eq:Boltzmann_eq_LHS}) 
and the collision term Eq.(\ref{eq:Boltzmann_eq_RHS}) 
into the Boltzmann equation Eq.(\ref{eq:Boltzmann_eq}) 
gives 
the following equation for $\Vec V_{\sigma}(\ve)$:  
\begin{equation}
\frac{\Vec F_{\sigma} (\ve)}{m}
=
\frac{\Vec V_{\sigma}(\ve)}{\tau_{\rm n}(\ve)} 
+
\frac{\sigma \Vec V_{\sigma}(\ve) \times \Vec e_z}{\tau_{\rm ss}(\ve)} 
+
\frac{2\alpha \sigma \Vec e_z \times \Vec F_{\sigma}(\ve)}{\tau_{\rm n}(\ve)}
. 
\end{equation}
Up to the first order of the spin-orbit coupling constant, $\alpha$, 
$\Vec V_{\sigma}(\ve)$ is obtained to be
\begin{eqnarray}
\Vec V_{\sigma}(\ve)
&&=
\frac{\tau_{\rm n}(\ve)}{m}
\left[
\Vec F_{\sigma} (\ve)
-\sigma \frac{\tau_{\rm n}(\ve)}{\tau_{\rm ss}(\ve)}
\Vec F_{\sigma} (\ve) \times \Vec e_z
\right]
\nonumber \\
&&+
2\alpha \sigma 
\Vec F_{\sigma} (\ve) \times \Vec e_z
.
\label{eq:V_sigma}
\end{eqnarray}
Substituting this formula of $\Vec V_{\sigma}(\ve)$ 
into that of $f_{\sigma}^{(1)}$ in Eq.(\ref{eq:f_1}), 
we obtain the distribution function, $f_{\sigma}(\Vec r, \Vec k)$, in Eq.(\ref{eq:f_sigma_k}) 
in the presence of the electric field, 
the chemical potential gradient, 
and the temperature gradient.

\subsection{Current densities and transport coefficients}
\label{eq:current_densities}

The number current density of spin-$\sigma$ electrons
is defined by
\begin{equation}
\Vec j^{\rm n\sigma}
=
\frac{1}{S} \sum_{i} \left<  {\Vec v}_i \right>_{\rm av}
\label{eq:jn0}
\end{equation}
where the summation is taken over spin-$\sigma$ electrons in the area $S$, and
$ {\Vec v}_i$ is the velocity operator of the $i$th electron 
given by 
\begin{equation}
 {\Vec v}_i
=
\frac{{\Vec p}_i}{m} 
+
2\alpha \sigma \Vec \nabla v_{\rm 2D}(\Vec r_i) \times \Vec e_z
. 
\end{equation}
The second term of ${\Vec v}_i$ comes from the spin-orbit interaction 
induced by the potential due to the electric field and impurities, $v_{\rm 2D}(\Vec r_i)$,  
and reduces to $- 2\alpha \sigma (d{\Vec p}_i/dt) \times \Vec e_z$ in $O(\alpha)$.  
The brackets in Eq.(\ref{eq:jn0}) 
take the average with respect to the wave packet in the steady state. 
In the steady state the acceleration by the electric field 
is balanced with the deceleration by the impurity potential 
when each wave packet travels through the system,  
that is $\left<d{\Vec p}_i/dt\right>_{\rm av}=0$, 
which leads to the vanishing contribution from
the second term of $ {\Vec v}_i$ to the current. 
This semiclassical argument made by Hankiewicz and Vignale\cite{Hankiewicz2006PRB} 
has been supported in terms of a rigorous density-matrix formalism by Culcer et al.\cite{Culcer2010} 
The first term of $ {\Vec v}_i$ gives 
\begin{equation}
\Vec j^{\rm n\sigma}
=
\frac{1}{S} \sum_{\Vec k} \frac{\hbar \Vec k}{m} 
\left(f^{(0)}(\ve_k) + f_{\sigma}^{(1)}(\Vec k) \right)
.
\end{equation}
Here the contribution from $f^{(0)}$ vanishes 
since $f^{(0)}$ depends only on the magnitude of $\Vec k$. 
Substituting the expression of $f_{\sigma}^{(1)}$, Eq.(\ref{eq:f_1}), 
we have 
\begin{equation}
\Vec j^{\rm n\sigma}
=
\langle
\rho \ve \Vec V_{\sigma}(\ve)
\rangle_{\sigma}
,
\label{eq:jn}
\end{equation}
where $\rho$ is the constant density of states per unit area per spin  
for two-dimensional electrons 
and 
the brackets represent the statistical average 
for spin-$\sigma$ electrons: 
\begin{equation}
\langle \cdots \rangle_{\sigma}
=\int_0^{\infty} d\ve \cdots 
\left( - \frac{\partial f^{(0)}(\ve, \mu_{\sigma}, \Te)}
{\partial \ve}
\right)
.
\end{equation}
The heat current density is obtained in a similar manner 
as 
\begin{equation}
\Vec j^{\rm q\sigma}
=
\langle
\rho \ve \Vec V_{\sigma}(\ve)(\ve-\mu_{\sigma})
\rangle_{\sigma}
.
\label{eq:jq}
\end{equation}

In the linear-response regime, 
each component of the number current density $\Vec j^{\rm n\sigma}$ 
is a linear function of components of thermodynamic forces, 
and the same is the case for $\Vec j^{\rm q\sigma}$. 
The thermodynamic force corresponding to each current density 
is obtained from the expression of the entropy production,\cite{Callen1960thermodynamics,Groot1962nonequilibrium} 
to be 
$-\Te^{-1}\Vec \nabla \mu_{\sigma}^{\rm ec}$ for $\Vec j^{\rm n\sigma}$ 
and $-\Te^{-2}\Vec \nabla \Te$ for $\Vec j^{\rm q\sigma}$. 
Therefore 
the linear relations between the current densities and the thermodynamic forces 
are written as
\begin{equation}
\left( \!\! \begin{array}{c}
                            \Vec j^{\rm n\sigma} \\ 
                             \Vec j^{\rm q\sigma} 
            \end{array}  \!\! \right)
 \!=\!      
\left( \!\! \begin{array}{cc}
                 L^{11\sigma}   & \!\!  L^{12\sigma}  \\ 
                 L^{21\sigma}   &  L^{22\sigma}     
           \end{array}  \!\! \right)   \!\! 
\left(  \!\! \begin{array}{c}
                           -\Vec \nabla \mu_{\sigma}^{\rm ec}           \\ 
                           -\Te^{-1} \Vec \nabla \Te   
            \end{array}    \!\! \right)  \!, 
\label{eq:jn_jq}
\end{equation}
with the transport coefficients 
\begin{equation}
L^{ij\sigma} \!= \!
\left( \!\! \begin{array}{rr}
                           L^{ij\sigma}_{xx}     &    L^{ij\sigma}_{xy}    \\ 
                           L^{ij\sigma}_{yx}     &    L^{ij\sigma}_{yy}    
           \end{array}  \!\! \right)   \! , \ \  
i=1,2,\ \ j=1,2. 
\end{equation}
The common factor $\Te^{-1}$ of the thermodynamic forces 
is absorbed in the transport coefficients.  
Since the 2DES in our model is isotropic in the $xy$ plane,
the transport coefficients have the following symmetry relation: 
$L^{ij\sigma}_{xx}=L^{ij\sigma}_{yy}$ and $L^{ij\sigma}_{xy}=-L^{ij\sigma}_{yx}$.

The expression for each transport coefficient 
is obtained by substituting the formula of $\Vec V_{\sigma}(\ve)$ 
in terms of the thermodynamic forces, 
Eq.(\ref{eq:V_sigma}) with Eq.(\ref{eq:generalized_force}), 
into those of the current densities, 
Eqs.(\ref{eq:jn}) and (\ref{eq:jq}). 
The obtained expression is
\begin{equation}
L^{ij\sigma}_{\mu \nu}
=
\langle L_{\mu \nu}(\ve) (\ve-\mu_{\sigma})^{i+j-2}\rangle_{\sigma}
, 
\end{equation}
with $\mu=x,y$ and $\nu=x,y$.  
Here $L_{\mu \nu}(\ve)$ is the contribution to the conductivity 
from electrons having energy $\ve$.  
Diagonal components 
\begin{equation}
L_{xx}(\ve)=L_{yy}(\ve)=
\rho \ve \frac{\tau_{\rm n}(\ve)}{m}
, 
\end{equation}
have the form of the Drude conductivity divided by $e^2$, 
while off-diagonal components 
\begin{equation}
L_{xy}(\ve)=-L_{yx}(\ve)=
\rho \ve \sigma 
\left[ - \frac{\tau_{\rm n}(\ve)^2}{m\tau_{\rm ss}(\ve)} +2\alpha\right]
,  
\end{equation}
are due to the spin-orbit interaction. 
The first term in the square brackets is the contribution from the skew scattering, 
while  the second term is that from the side jump. 
Note that the spin-orbit interaction gives rise to 
all off-diagonal transport coefficients 
in $L^{11\sigma}$, $L^{12\sigma}$, $L^{21\sigma}$, and $L^{22\sigma}$. 

When the 2DES is degenerate ($\mu_{\sigma} \gg \kB \Te$), 
\begin{equation}
L^{11\sigma}_{\mu \nu}=
L_{\mu \nu}(\mu_{\sigma})
, \ \ 
L^{12\sigma}_{\mu \nu}
=
\frac{\pi^2}{3} (\kB \Te)^2 
\left[ \frac{dL_{\mu \nu}(\ve)}{d\ve}\right]_{\ve=\mu_{\sigma}}. 
\label{eq:L_degenerate}
\end{equation}
Therefore the Mott relation\cite{Mott-Jones} holds, 
that is, the thermoelectric conductivity tensor, $L^{12\sigma}_{\mu \nu}$, is proportional to 
the energy derivative of the electric conductivity tensor, $L^{11\sigma}_{\mu \nu} $. 
In addition, 
the diagonal electric conductivity reduces to the Drude conductivity, 
$e^2 L^{11\sigma}_{xx} = n_{\sigma} e^2 \tau_{\rm n}(\mu_{\sigma})/m$, 
where $n_{\sigma}$ is the density of spin-$\sigma$ electrons.

In the discussion of the spin Nernst effect as well as the spin Hall effect, 
it is convenient to reorganize the number and heat current densities for both spins into  
the spin current density, $\Vec j^{\rm s}$, 
the number current density $\Vec j^{\rm n}$, 
and
the heat current density, $\Vec j^{\rm q}$, 
as follows, 
\begin{equation}
\begin{split}
\Vec j^{\rm s} &= (\Vec j^{\rm n\uparrow} - \Vec j^{\rm n\downarrow})/2, \\ 
\Vec j^{\rm n} &= \Vec j^{\rm n\uparrow} + \Vec j^{\rm n\downarrow}, \\  
\Vec j^{\rm q} &= \Vec j^{\rm q\uparrow} + \Vec j^{\rm q\downarrow}
, 
\end{split}
\end{equation}
where we have used the notation $\sigma= \uparrow, \downarrow $ 
instead of $\sigma=+1,-1$. 
The corresponding 
thermodynamic forces\cite{Callen1960thermodynamics,Groot1962nonequilibrium} are 
$-\Te^{-1}\Vec \nabla \mu^{\rm s}_{\rm ec}$, 
$-\Te^{-1}\Vec \nabla \mu^{\rm n}_{\rm ec}$, 
and $-\Te^{-2} \Vec \nabla \Te$, respectively, with 
\begin{equation}
\begin{split}
\mu^{\rm s}_{\rm ec} &= \mu_{\uparrow}^{\rm ec} - \mu_{\downarrow}^{\rm ec}
=\mu_{\uparrow} - \mu_{\downarrow},
\\  
\mu^{\rm n}_{\rm ec} &= (\mu_{\uparrow}^{\rm ec} + \mu_{\downarrow}^{\rm ec})/2 
. 
\end{split}
\end{equation}
The linear relations now become 
\begin{equation}
\left( \!\! \begin{array}{c}
                            \Vec j^{\rm s} \\ 
                            \Vec j^{\rm n} \\ 
                             \Vec j^{\rm q} 
            \end{array}  \!\! \right)
 \!=\!      
\left( \!\! \begin{array}{ccc}
                 L^{\rm ss}   & \!\!  L^{\rm sn}  & \!\!  L^{\rm sq}\\ 
                 L^{\rm ns}   & \!\!  L^{\rm nn}  & \!\!  L^{\rm nq}\\ 
                 L^{\rm qs}   & \!\!  L^{\rm qn}  & \!\!  L^{\rm qq}
           \end{array}  \!\! \right)   \!\! 
\left(  \!\! \begin{array}{c}
                            -\Vec \nabla \mu^{\rm s}_{\rm ec}           \\ 
                           -\Vec \nabla \mu^{\rm n}_{\rm ec}           \\ 
                          -\Te^{-1} \Vec \nabla \Te   
            \end{array}    \!\! \right)  \!, 
\label{eq:js_jn_jq}
\end{equation}
where 
\begin{eqnarray}
L^{\rm ss}& \!=\! &\frac{L^{11\uparrow} \!+\! L^{11\downarrow}}{4},  \ 
L^{\rm sn} \!=\! \frac{L^{11\uparrow} \!-\! L^{11\downarrow}}{2}, \  
L^{\rm sq} \!=\! \frac{L^{21\uparrow} \!-\! L^{21\downarrow}}{2},\nonumber\\ 
L^{\rm ns}& \!=\! &L^{\rm sn},  \ \ 
L^{\rm nn} \!=\! L^{11\uparrow} \!+\! L^{11\downarrow}, \ \ 
L^{\rm nq} \!=\! L^{21\uparrow} \!+\! L^{21\downarrow}, \\ 
L^{\rm qs}& \!=\! & L^{\rm sq}, \ \ 
L^{\rm qn} \!=\! L^{\rm nq},   \ \ \ \ \ \ \ \ \ \ \ 
L^{\rm qq} \!=\! L^{22\uparrow} \!+\! L^{22\downarrow}. \nonumber
\end{eqnarray}

In the following 
we employ the condition satisfied in nonmagnetic systems, that is,
the chemical potentials for both spins are the same in equilibrium, 
$\mu_{\uparrow}^{(0)}=\mu_{\downarrow}^{(0)}$.    
Then we have $L^{ij\uparrow}_{xx}=L^{ij\downarrow}_{xx}$ and $L^{ij\uparrow}_{xy}=-L^{ij\downarrow}_{xy}$. 
With use of these relations, we confirm that 
the Onsager relation\cite{Onsager1931_I, Onsager1931_II} is satisfied, that is 
$L^{ij\uparrow}_{\mu \nu}=L^{ji\downarrow}_{\nu \mu}$.  
In addition we have 
\begin{equation}
\begin{split}
L^{ij\uparrow}+L^{ij\downarrow} &= 2 L^{ij\uparrow}_{xx} I, \ \ 
I=\left( \!\! \begin{array}{rr}
                           1     &    0    \\ 
                           0     &    1   
           \end{array}  \!\! \right) , \\
L^{ij\uparrow}-L^{ij\downarrow} &= 2 L^{ij\uparrow}_{xy} J, \ \ 
J=\left( \!\! \begin{array}{rr}
                           0     &    1    \\ 
                           -1     &    0   
           \end{array}  \!\! \right) ,
\label{eq:diagonal_offdiagonal}
\end{split}
\end{equation}
from which we find that $L^{\rm sn}$($=L^{\rm ns}$) and $L^{\rm sq}$($=L^{\rm qs}$) 
are proportional to $J$, 
while the other matrices in Eq.(\ref{eq:js_jn_jq}) are proportional to $I$. 
Therefore we can separate Eq.(\ref{eq:js_jn_jq}) representing the linear relations into 
the following two equations: 
\begin{equation}
\left( \!\! \begin{array}{c}
                            j^{\rm s}_x \\ 
                            j^{\rm n}_y \\ 
                             j^{\rm q}_y 
            \end{array}  \!\! \right)
 \!=\!      
\left( \!\! \begin{array}{ccc}
                 L^{\rm ss}_{xx}   & \!\!  L^{\rm sn}_{xy}  & \!\!  L^{\rm sq}_{xy}\\ 
                 L^{\rm ns}_{yx}   & \!\!  L^{\rm nn}_{yy}  & \!\!  L^{\rm nq}_{yy}\\ 
                 L^{\rm qs}_{yx}   & \!\!  L^{\rm qn}_{yy}  & \!\!  L^{\rm qq}_{yy}
           \end{array}  \!\! \right)   \!\! 
\left(  \!\! \begin{array}{c}
                            -\nabla_x \mu^{\rm s}_{\rm ec}           \\ 
                           -\nabla_y \mu^{\rm n}_{\rm ec}           \\ 
                          -\Te^{-1} \nabla_y \Te   
            \end{array}    \!\! \right)  \!, 
\label{eq:js_jn_jq_a}
\end{equation}
and
\begin{equation}
\left( \!\! \begin{array}{c}
                            j^{\rm s}_y \\ 
                            j^{\rm n}_x \\ 
                             j^{\rm q}_x 
            \end{array}  \!\! \right)
 \!=\!      
\left( \!\! \begin{array}{ccc}
                 L^{\rm ss}_{yy}   & \!\!  L^{\rm sn}_{yx}  & \!\!  L^{\rm sq}_{yx}\\ 
                 L^{\rm ns}_{xy}   & \!\!  L^{\rm nn}_{xx}  & \!\!  L^{\rm nq}_{xx}\\ 
                 L^{\rm qs}_{xy}   & \!\!  L^{\rm qn}_{xx}  & \!\!  L^{\rm qq}_{xx}
           \end{array}  \!\! \right)   \!\! 
\left(  \!\! \begin{array}{c}
                            -\nabla_y \mu^{\rm s}_{\rm ec}           \\ 
                           -\nabla_x \mu^{\rm n}_{\rm ec}           \\ 
                          -\Te^{-1} \nabla_x \Te   
            \end{array}    \!\! \right)  \!. 
\label{eq:js_jn_jq_b}
\end{equation}
These equations indicate that, in nonmagnetic systems,  
the spin current (say, along the $x$ axis) 
is coupled only to the perpendicular component of the number and heat currents
(along the $y$ axis). 

\section{spin Nernst effect}
\label{sec:spin_Nernst}
\subsection{Calculation of the spin Nernst coefficient}

We consider a state 
in which all current densities are uniform in a rectangular sample in the $xy$ plane. 
In this state the thermodynamic forces are also uniform 
as derived from Eqs.(\ref{eq:js_jn_jq_a}) and (\ref{eq:js_jn_jq_b}). 
We apply a uniform temperature gradient along the $x$ axis 
($\nabla_x \Te = {\rm const.} \not=0$, $\nabla_y \Te=0$), 
under the condition that both the number current and the spin current are vanishing 
($\Vec j^{\rm s}=0$ and $\Vec j^{\rm n}=0$). 
The spin Nernst effect in the absence of the spin relaxation is 
the appearance of a uniform gradient along the $y$ axis of 
the chemical-potential difference between up and down spins, 
$\mu^{\rm s}_{\rm ec}=\mu_{\uparrow}-\mu_{\downarrow}$,  
proportional to the applied temperature gradient along the $x$ axis: 
\begin{equation}
\nabla_y \mu^{\rm s}_{\rm ec} 
=
N_{\rm s} \nabla_x \Te
.
\end{equation}
Here we call $N_{\rm s}$ the spin Nernst coefficient. 

To obtain the formula of $N_{\rm s}$ in terms of transport coefficients, 
we write the conditions of $j^{\rm s}_y=0$ and $j^{\rm n}_x=0$  
in terms of the thermodynamic forces  
using Eq.(\ref{eq:js_jn_jq_b}) 
and eliminate $\nabla_x \mu^{\rm n}_{\rm ec}$. 
Then we obtain 
\begin{equation}
N_{\rm s} 
= 
-\frac{1}{\Te}  
\frac
{L^{\rm nn}_{xx}L^{\rm sq}_{yx} - L^{\rm sn}_{yx}L^{\rm nq}_{xx}}
{L^{\rm nn}_{xx}L^{\rm ss}_{yy} - L^{\rm sn}_{yx}L^{\rm ns}_{xy}}
,
\end{equation}
which becomes, in the first order of $\alpha$, 
\begin{equation}
N_{\rm s} 
= 
\frac{2}{\Te}  
\frac
{L^{11\uparrow}_{xx}L^{21\uparrow}_{xy} - L^{11\uparrow}_{xy}L^{21\uparrow}_{xx}}
{\left(L^{11\uparrow}_{xx}\right)^{2}}
.
\end{equation}
On the other hand, Eq.(\ref{eq:js_jn_jq_a}) with 
$j^{\rm s}_x=0$, $j^{\rm n}_y=0$, and $\nabla_y \Te=0$ 
gives 
$\nabla_x \mu^{\rm s}_{\rm ec}=0$, 
$\nabla_y \mu^{\rm n}_{\rm ec} =0$,
and  
$j_y^{\rm q}=0$.
In particular $\nabla_x \mu^{\rm s}_{\rm ec}=0$ 
means that no spin accumulation is generated 
in the same direction as the applied temperature gradient. 

In calculating the spin Nernst coefficient, 
we consider the degenerate electron gas 
in which the equilibrium chemical potential, 
$\mu=\mu_{\uparrow}^{(0)}=\mu_{\downarrow}^{(0)}$, 
is much larger than $\kB \Te$. 
In this case, using Eq.(\ref{eq:L_degenerate}), we obtain
\begin{equation}
N_{\rm s} 
= 
\frac{2\pi^2 \kB^2 \Te}{3} 
\left[ 
\frac{\tau_{\rm n}'(\mu)}{\tau_{\rm n}(\mu)} 
n_{\rm s}
+ 
\frac{\tau_{\rm ss}'(\mu)}{\tau_{\rm ss}(\mu)} 
\frac{\tau_{\rm n}(\mu)}{\tau_{\rm ss}(\mu)}
\right]
,
\label{eq:N_s}
\end{equation}
with
\begin{equation}
n_{\rm s}
=
- \frac{\tau_{\rm n}(\mu)}{\tau_{\rm ss}(\mu)}
- \frac{2m\alpha}{\tau_{\rm n}(\mu)}
,
\label{eq:ns}
\end{equation}
and
\begin{equation}
\tau_{\rm n}'(\mu) = \left[ \frac{d \tau_{\rm n}(\ve)}{d\ve}\right]_{\ve=\mu} .
\end{equation}
The first term of $n_{\rm s}$ in Eq.(\ref{eq:ns}) comes from the skew scattering, 
while the second term is from the side jump. 
Hankiewicz and Vignale\cite{Hankiewicz2009} have shown, 
in the calculation for a model impurity potential, 
that 
the energy dependence of $\tau_{\rm ss}$ is smaller than 
that of $\tau_{\rm n}$ at the Fermi energy. 
Therefore the term with $n_{\rm s}$ is dominant
in $N_{\rm s}$ in Eq.(\ref{eq:N_s}).  
They have also shown that 
$\tau_{\rm ss}$ is negative (positive) 
for repulsive (attractive) impurity potentials.\cite{Hankiewicz2006PRB} 
On the other hand, the spin-orbit coupling constant, $\alpha$, is positive for semiconductors. 

\begin{figure}[ht]
\includegraphics[height=9cm, bb=0 0 420 595]{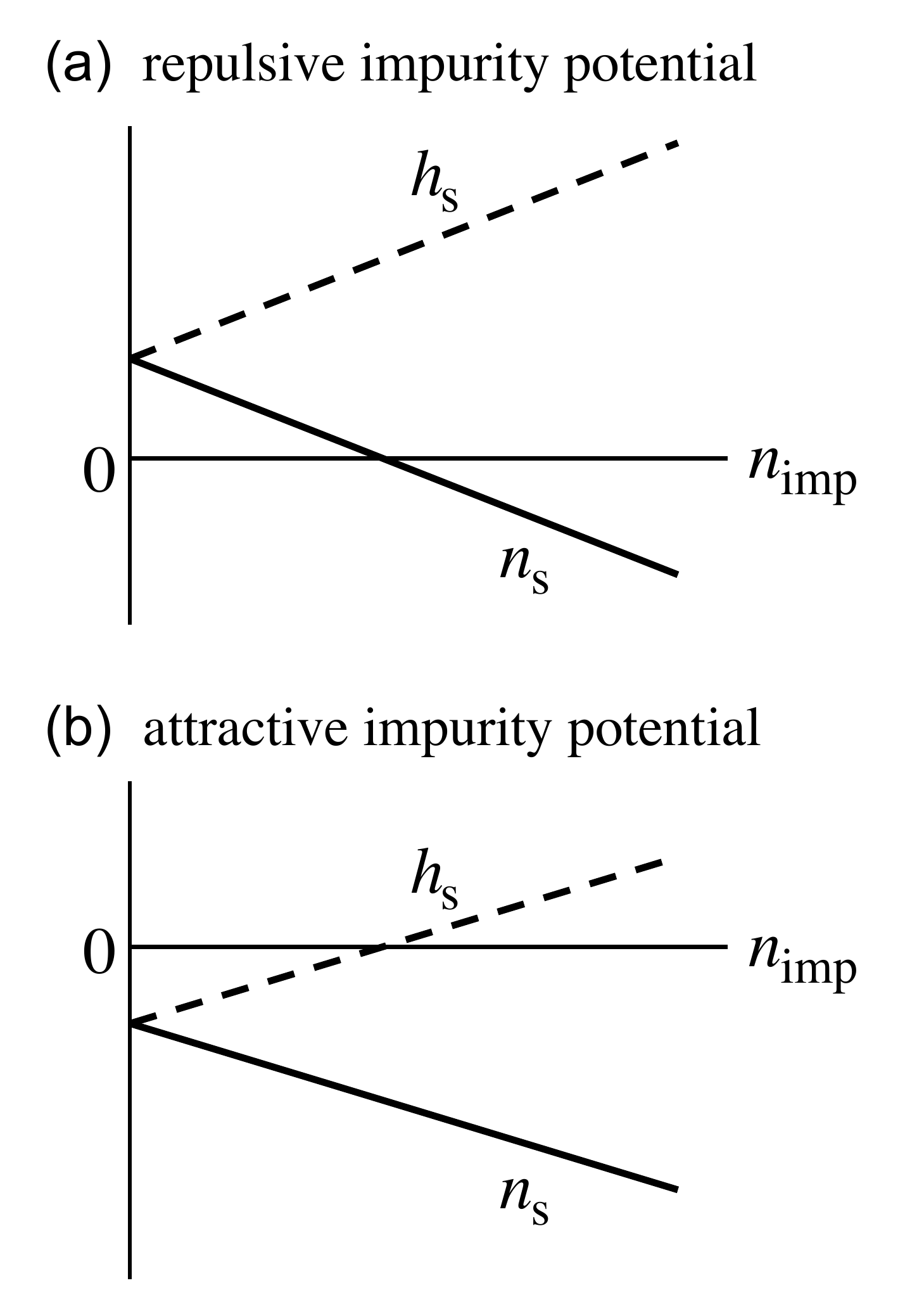}
\caption{\label{fig:ns_hs}
Normalized spin Nernst coefficient, $n_{\rm s}$ in Eq.(\ref{eq:ns}) 
and the normalized spin Hall coefficient, $h_{\rm s}$ in Eq.(\ref{eq:hs}) , 
as a function of the impurity density, $n_{\rm imp}$,  
for (a) a repulsive impurity potential and 
for (b) an attractive impurity potential.}
\end{figure}

According to Kohn and Luttinger,\cite{Kohn-Luttinger1957} 
both $1/\tau_{\rm n}$ and $1/\tau_{\rm ss}$ 
are proportional to the impurity density, $n_{\rm imp}$, 
up to the third order 
in the expansion with respect to the strength of the impurity potential.  
Therefore we employ this proportionality by considering weak impurity potentials. 
Then in Eq.(\ref{eq:ns}) the first term of $n_{\rm s}$ from the skew scattering is 
independent of $n_{\rm imp}$, 
while the second term from the side jump 
is linear in $n_{\rm imp}$ and the coefficient is negative. 
In the case of the repulsive impurity potential 
where the first term of $n_{\rm s}$ is positive, 
$n_{\rm s}$ changes its sign when $n_{\rm imp}$ is increased, 
as shown in Fig.\ref{fig:ns_hs}, 
while, for the attractive impurity potential, $n_{\rm s}$ is negative at any value of $n_{\rm imp}$. 

\subsection{Comparison with the spin Hall coefficient}

We compare the dependence of the spin Nernst coefficient 
on the impurity density with that of the spin Hall coefficient 
derived in Ref.\onlinecite{Hankiewicz2006PRB}.
We apply the number current along the $x$ axis, 
while we keep the spin current vanishing. 
We also set the condition that 
the current along the $y$ axis is vanishing for both the number and the spin,  
and that the electron temperature is uniform. 
The condition of $j_y^{\rm s}=0$ with $\nabla_x \Te=0$ in Eq.(\ref{eq:js_jn_jq_b})
gives immediately
\begin{equation}
\nabla_y \mu^{\rm s}_{\rm ec} 
=
H_{\rm s} \nabla_x \mu^{\rm n}_{\rm ec}
, 
\end{equation}
with the spin Hall coefficient $H_{\rm s}$ given by
\begin{equation}
H_{\rm s} 
= 
 - \frac{L^{\rm sn}_{yx}}{L^{\rm ss}_{yy}}
= 
 2\ \frac{L_{xy}^{11 \uparrow}}{L_{xx}^{11 \uparrow}}
.
\end{equation}

When the electron gas is degenerate, we obtain 
\begin{equation}
H_{\rm s} = 2 h_{\rm s}, \ \ 
h_{\rm s} = 
- \frac{\tau_{\rm n}(\mu)}{\tau_{\rm ss}(\mu)}
+ \frac{2m\alpha}{\tau_{\rm n}(\mu)}
,
\label{eq:hs}
\end{equation}
which reproduces 
the dependence of the spin Hall conductivity on $\tau_{\rm n}$ and $\tau_{\rm ss}$ 
derived in 2DES by Hankiewicz and Vignale.\cite{Hankiewicz2006PRB} 
Comparing this formula of $h_{\rm s}$ with that of $n_{\rm s}$ in Eq.(\ref{eq:ns}), 
the difference appears 
only at the sign of the second term from the side jump: 
the slope of $h_{\rm s}$ as a function of $n_{\rm imp}$ is positive, 
while that of $n_{\rm s}$ is negative.
Therefore the change in sign of $h_{\rm s}$ with $n_{\rm imp}$ 
appears for the attractive impurity potential as shown in Fig.\ref{fig:ns_hs}.

The sample in the experiment by Sih et al.,\cite{Sih2005} 
in which Si donors are doped in the quantum well, 
corresponds to the attractive impurity potential in Fig.\ref{fig:ns_hs}(b). 
According to the calculation for this sample by 
Hankiewicz and Vignale,\cite{Hankiewicz2006PRB}
the contribution to the spin Hall conductivity  
from the side jump is comparable in size to 
that from the skew scattering. 
Therefore we expect that the sign change of the spin Hall coefficient should 
be observed if the density of Si impurities in the well 
is changed around the value of the Si density 
used in the experiment. 

A Be impurity in GaAs is known to act as an acceptor. 
Therefore 
doping Be in a quantum well introduces 
the repulsive impurity potential for the 2DES.\cite{Haug1987}  
In this case 
it is expected, according to the calculated result in Fig.\ref{fig:ns_hs}(a), that
the spin Nernst coefficient changes its sign 
as a function of the density of Be impurities.

\section{Conclusions}
\label{eq:conclusions}

We have studied theoretically 
the spin Nernst effect due to the spin-orbit interaction in the extrinsic origin 
in two-dimensional electron systems (2DES) in the $xy$ plane
by employing the Boltzmann equation.  
We consider a 2DES confined within a symmetric quantum well 
with delta doping at the center of the well.    
We have shown in such a 2DES that 
the spin-orbit interaction, including that induced by the impurity potential, 
is diagonal in the $z$ component of spin because of the symmetry of the system. 

In this model of 2DES we have investigated the dependence of the spin Nernst coefficient 
on the sign of the impurity potential and on the impurity density, 
and compared the result with that of the spin Hall coefficient. 
We have found that the spin Nernst coefficient changes its sign 
as a function of the impurity density 
in the case of the repulsive impurity potential, 
while no sign change occurs for the attractive impurity potential.  
On the other hand, 
the spin Hall coefficient changes its sign 
in the case of the attractive impurity potential,
as shown already by Hankiewicz and Vignale.\cite{Hankiewicz2006PRB} 
The sign change of each coefficient occurs due to 
the cancellation between 
the skew-scattering contribution and the side-jump contribution. 


%


\begin{thebibliography}{51}%
\makeatletter
\providecommand \@ifxundefined [1]{%
 \@ifx{#1\undefined}
}%
\providecommand \@ifnum [1]{%
 \ifnum #1\expandafter \@firstoftwo
 \else \expandafter \@secondoftwo
 \fi
}%
\providecommand \@ifx [1]{%
 \ifx #1\expandafter \@firstoftwo
 \else \expandafter \@secondoftwo
 \fi
}%
\providecommand \natexlab [1]{#1}%
\providecommand \enquote  [1]{``#1''}%
\providecommand \bibnamefont  [1]{#1}%
\providecommand \bibfnamefont [1]{#1}%
\providecommand \citenamefont [1]{#1}%
\providecommand \href@noop [0]{\@secondoftwo}%
\providecommand \href [0]{\begingroup \@sanitize@url \@href}%
\providecommand \@href[1]{\@@startlink{#1}\@@href}%
\providecommand \@@href[1]{\endgroup#1\@@endlink}%
\providecommand \@sanitize@url [0]{\catcode `\\12\catcode `\$12\catcode
  `\&12\catcode `\#12\catcode `\^12\catcode `\_12\catcode `\%12\relax}%
\providecommand \@@startlink[1]{}%
\providecommand \@@endlink[0]{}%
\providecommand \url  [0]{\begingroup\@sanitize@url \@url }%
\providecommand \@url [1]{\endgroup\@href {#1}{\urlprefix }}%
\providecommand \urlprefix  [0]{URL }%
\providecommand \Eprint [0]{\href }%
\providecommand \doibase [0]{http://dx.doi.org/}%
\providecommand \selectlanguage [0]{\@gobble}%
\providecommand \bibinfo  [0]{\@secondoftwo}%
\providecommand \bibfield  [0]{\@secondoftwo}%
\providecommand \translation [1]{[#1]}%
\providecommand \BibitemOpen [0]{}%
\providecommand \bibitemStop [0]{}%
\providecommand \bibitemNoStop [0]{.\EOS\space}%
\providecommand \EOS [0]{\spacefactor3000\relax}%
\providecommand \BibitemShut  [1]{\csname bibitem#1\endcsname}%
\let\auto@bib@innerbib\@empty
\bibitem [{\citenamefont {Sinova}\ \emph {et~al.}(2006)\citenamefont {Sinova},
  \citenamefont {Murakami}, \citenamefont {Shen},\ and\ \citenamefont
  {Choi}}]{Sinova_et_al_2006}%
  \BibitemOpen
  \bibfield  {author} {\bibinfo {author} {\bibfnamefont {J.}~\bibnamefont
  {Sinova}}, \bibinfo {author} {\bibfnamefont {S.}~\bibnamefont {Murakami}},
  \bibinfo {author} {\bibfnamefont {S.~Q.}\ \bibnamefont {Shen}}, \ and\
  \bibinfo {author} {\bibfnamefont {M.~S.}\ \bibnamefont {Choi}},\ }\href@noop
  {} {\bibfield  {journal} {\bibinfo  {journal} {Solid State Commun.}\ }\textbf
  {\bibinfo {volume} {138}},\ \bibinfo {pages} {214} (\bibinfo {year}
  {2006})}\BibitemShut {NoStop}%
\bibitem [{\citenamefont {Dyakonov}\ and\ \citenamefont
  {Khaetskii}(2008)}]{Dyakonov2008}%
  \BibitemOpen
  \bibfield  {author} {\bibinfo {author} {\bibfnamefont {M.~I.}\ \bibnamefont
  {Dyakonov}}\ and\ \bibinfo {author} {\bibfnamefont {A.~V.}\ \bibnamefont
  {Khaetskii}},\ }in\ \href {http://dx.doi.org/10.1007/978-3-540-78820-1_8}
  {\emph {\bibinfo {booktitle} {Spin Physics in Semiconductors}}},\ \bibinfo
  {editor} {edited by\ \bibinfo {editor} {\bibfnamefont {M.~I.}\ \bibnamefont
  {Dyakonov}}}\ (\bibinfo  {publisher} {Springer Berlin Heidelberg},\ \bibinfo
  {year} {2008})\ pp.\ \bibinfo {pages} {211--243}\BibitemShut {NoStop}%
\bibitem [{\citenamefont {Vignale}(2010)}]{Vignale2010}%
  \BibitemOpen
  \bibfield  {author} {\bibinfo {author} {\bibfnamefont {G.}~\bibnamefont
  {Vignale}},\ }\href@noop {} {\bibfield  {journal} {\bibinfo  {journal} {J.
  Supercond. Nov. Magn.}\ }\textbf {\bibinfo {volume} {23}},\ \bibinfo {pages}
  {3} (\bibinfo {year} {2010})}\BibitemShut {NoStop}%
\bibitem [{\citenamefont {{\v{Z}}uti{\'c}}\ \emph {et~al.}(2004)\citenamefont
  {{\v{Z}}uti{\'c}}, \citenamefont {Fabian},\ and\ \citenamefont
  {Sarma}}]{Zutic2004spintronics}%
  \BibitemOpen
  \bibfield  {author} {\bibinfo {author} {\bibfnamefont {I.}~\bibnamefont
  {{\v{Z}}uti{\'c}}}, \bibinfo {author} {\bibfnamefont {J.}~\bibnamefont
  {Fabian}}, \ and\ \bibinfo {author} {\bibfnamefont {S.~D.}\ \bibnamefont
  {Sarma}},\ }\href@noop {} {\bibfield  {journal} {\bibinfo  {journal} {Rev.
  Mod. Phys.}\ }\textbf {\bibinfo {volume} {76}},\ \bibinfo {pages} {323}
  (\bibinfo {year} {2004})}\BibitemShut {NoStop}%
\bibitem [{\citenamefont {Kato}\ \emph {et~al.}(2004)\citenamefont {Kato},
  \citenamefont {Myers}, \citenamefont {Gossard},\ and\ \citenamefont
  {Awschalom}}]{Kato2004}%
  \BibitemOpen
  \bibfield  {author} {\bibinfo {author} {\bibfnamefont {Y.~K.}\ \bibnamefont
  {Kato}}, \bibinfo {author} {\bibfnamefont {R.~C.}\ \bibnamefont {Myers}},
  \bibinfo {author} {\bibfnamefont {A.~C.}\ \bibnamefont {Gossard}}, \ and\
  \bibinfo {author} {\bibfnamefont {D.~D.}\ \bibnamefont {Awschalom}},\
  }\href@noop {} {\bibfield  {journal} {\bibinfo  {journal} {Science}\ }\textbf
  {\bibinfo {volume} {306}},\ \bibinfo {pages} {1910} (\bibinfo {year}
  {2004})}\BibitemShut {NoStop}%
\bibitem [{\citenamefont {Wunderlich}\ \emph {et~al.}(2005)\citenamefont
  {Wunderlich}, \citenamefont {Kaestner}, \citenamefont {Sinova},\ and\
  \citenamefont {Jungwirth}}]{Wunderlich2005}%
  \BibitemOpen
  \bibfield  {author} {\bibinfo {author} {\bibfnamefont {J.}~\bibnamefont
  {Wunderlich}}, \bibinfo {author} {\bibfnamefont {B.}~\bibnamefont
  {Kaestner}}, \bibinfo {author} {\bibfnamefont {J.}~\bibnamefont {Sinova}}, \
  and\ \bibinfo {author} {\bibfnamefont {T.}~\bibnamefont {Jungwirth}},\
  }\href@noop {} {\bibfield  {journal} {\bibinfo  {journal} {Phys. Rev. Lett.}\
  }\textbf {\bibinfo {volume} {94}},\ \bibinfo {pages} {47204} (\bibinfo {year}
  {2005})}\BibitemShut {NoStop}%
\bibitem [{\citenamefont {Sih}\ \emph {et~al.}(2005)\citenamefont {Sih},
  \citenamefont {Myers}, \citenamefont {Kato}, \citenamefont {Lau},
  \citenamefont {Gossard},\ and\ \citenamefont {Awschalom}}]{Sih2005}%
  \BibitemOpen
  \bibfield  {author} {\bibinfo {author} {\bibfnamefont {V.}~\bibnamefont
  {Sih}}, \bibinfo {author} {\bibfnamefont {R.~C.}\ \bibnamefont {Myers}},
  \bibinfo {author} {\bibfnamefont {Y.~K.}\ \bibnamefont {Kato}}, \bibinfo
  {author} {\bibfnamefont {W.~H.}\ \bibnamefont {Lau}}, \bibinfo {author}
  {\bibfnamefont {A.~C.}\ \bibnamefont {Gossard}}, \ and\ \bibinfo {author}
  {\bibfnamefont {D.~D.}\ \bibnamefont {Awschalom}},\ }\href@noop {} {\bibfield
   {journal} {\bibinfo  {journal} {Nat. Phys.}\ }\textbf {\bibinfo {volume}
  {1}},\ \bibinfo {pages} {31} (\bibinfo {year} {2005})}\BibitemShut {NoStop}%
\bibitem [{\citenamefont {Murakami}\ \emph {et~al.}(2003)\citenamefont
  {Murakami}, \citenamefont {Nagaosa},\ and\ \citenamefont
  {Zhang}}]{Murakami2003}%
  \BibitemOpen
  \bibfield  {author} {\bibinfo {author} {\bibfnamefont {S.}~\bibnamefont
  {Murakami}}, \bibinfo {author} {\bibfnamefont {N.}~\bibnamefont {Nagaosa}}, \
  and\ \bibinfo {author} {\bibfnamefont {S.-C.}\ \bibnamefont {Zhang}},\ }\href
  {\doibase 10.1126/science.1087128} {\bibfield  {journal} {\bibinfo  {journal}
  {Science}\ }\textbf {\bibinfo {volume} {301}},\ \bibinfo {pages} {1348}
  (\bibinfo {year} {2003})}\BibitemShut {NoStop}%
\bibitem [{\citenamefont {Sinova}\ \emph {et~al.}(2004)\citenamefont {Sinova},
  \citenamefont {Culcer}, \citenamefont {Niu}, \citenamefont {Sinitsyn},
  \citenamefont {Jungwirth},\ and\ \citenamefont {MacDonald}}]{Sinova2004}%
  \BibitemOpen
  \bibfield  {author} {\bibinfo {author} {\bibfnamefont {J.}~\bibnamefont
  {Sinova}}, \bibinfo {author} {\bibfnamefont {D.}~\bibnamefont {Culcer}},
  \bibinfo {author} {\bibfnamefont {Q.}~\bibnamefont {Niu}}, \bibinfo {author}
  {\bibfnamefont {N.~A.}\ \bibnamefont {Sinitsyn}}, \bibinfo {author}
  {\bibfnamefont {T.}~\bibnamefont {Jungwirth}}, \ and\ \bibinfo {author}
  {\bibfnamefont {A.~H.}\ \bibnamefont {MacDonald}},\ }\href {\doibase
  10.1103/PhysRevLett.92.126603} {\bibfield  {journal} {\bibinfo  {journal}
  {Phys. Rev. Lett.}\ }\textbf {\bibinfo {volume} {92}},\ \bibinfo {pages}
  {126603} (\bibinfo {year} {2004})}\BibitemShut {NoStop}%
\bibitem [{\citenamefont {Dyakonov}\ and\ \citenamefont
  {Perel}(1971{\natexlab{a}})}]{Dyakonov1971JETPL}%
  \BibitemOpen
  \bibfield  {author} {\bibinfo {author} {\bibfnamefont {M.~I.}\ \bibnamefont
  {Dyakonov}}\ and\ \bibinfo {author} {\bibfnamefont {V.~I.}\ \bibnamefont
  {Perel}},\ }\href@noop {} {\bibfield  {journal} {\bibinfo  {journal} {Sov.
  Phys. JETP}\ }\textbf {\bibinfo {volume} {13}},\ \bibinfo {pages} {467}
  (\bibinfo {year} {1971}{\natexlab{a}})}\BibitemShut {NoStop}%
\bibitem [{\citenamefont {Dyakonov}\ and\ \citenamefont
  {Perel}(1971{\natexlab{b}})}]{Dyakonov1971PhysicsLettersA}%
  \BibitemOpen
  \bibfield  {author} {\bibinfo {author} {\bibfnamefont {M.~I.}\ \bibnamefont
  {Dyakonov}}\ and\ \bibinfo {author} {\bibfnamefont {V.~I.}\ \bibnamefont
  {Perel}},\ }\href {\doibase 10.1016/0375-9601(71)90196-4} {\bibfield
  {journal} {\bibinfo  {journal} {Phys. Lett. A}\ }\textbf {\bibinfo {volume}
  {35}},\ \bibinfo {pages} {459 } (\bibinfo {year}
  {1971}{\natexlab{b}})}\BibitemShut {NoStop}%
\bibitem [{\citenamefont {Hirsch}(1999)}]{Hirsch1999}%
  \BibitemOpen
  \bibfield  {author} {\bibinfo {author} {\bibfnamefont {J.~E.}\ \bibnamefont
  {Hirsch}},\ }\href@noop {} {\bibfield  {journal} {\bibinfo  {journal} {Phys.
  Rev. Lett.}\ }\textbf {\bibinfo {volume} {83}},\ \bibinfo {pages} {1834}
  (\bibinfo {year} {1999})}\BibitemShut {NoStop}%
\bibitem [{\citenamefont {Zhang}(2000)}]{Zhang2000}%
  \BibitemOpen
  \bibfield  {author} {\bibinfo {author} {\bibfnamefont {S.}~\bibnamefont
  {Zhang}},\ }\href@noop {} {\bibfield  {journal} {\bibinfo  {journal} {Phys.
  Rev. Lett.}\ }\textbf {\bibinfo {volume} {85}},\ \bibinfo {pages} {393}
  (\bibinfo {year} {2000})}\BibitemShut {NoStop}%
\bibitem [{\citenamefont {Engel}\ \emph {et~al.}(2005)\citenamefont {Engel},
  \citenamefont {Halperin},\ and\ \citenamefont {Rashba}}]{Engel2005}%
  \BibitemOpen
  \bibfield  {author} {\bibinfo {author} {\bibfnamefont {H.-A.}\ \bibnamefont
  {Engel}}, \bibinfo {author} {\bibfnamefont {B.~I.}\ \bibnamefont {Halperin}},
  \ and\ \bibinfo {author} {\bibfnamefont {E.~I.}\ \bibnamefont {Rashba}},\
  }\href {\doibase 10.1103/PhysRevLett.95.166605} {\bibfield  {journal}
  {\bibinfo  {journal} {Phys. Rev. Lett.}\ }\textbf {\bibinfo {volume} {95}},\
  \bibinfo {pages} {166605} (\bibinfo {year} {2005})}\BibitemShut {NoStop}%
\bibitem [{\citenamefont {Hankiewicz}\ and\ \citenamefont
  {Vignale}(2006)}]{Hankiewicz2006PRB}%
  \BibitemOpen
  \bibfield  {author} {\bibinfo {author} {\bibfnamefont {E.~M.}\ \bibnamefont
  {Hankiewicz}}\ and\ \bibinfo {author} {\bibfnamefont {G.}~\bibnamefont
  {Vignale}},\ }\href {\doibase 10.1103/PhysRevB.73.115339} {\bibfield
  {journal} {\bibinfo  {journal} {Phys. Rev. B}\ }\textbf {\bibinfo {volume}
  {73}},\ \bibinfo {pages} {115339} (\bibinfo {year} {2006})}\BibitemShut
  {NoStop}%
\bibitem [{\citenamefont {Tse}\ and\ \citenamefont
  {Das~Sarma}(2006)}]{Tse-Das_Sarma2006}%
  \BibitemOpen
  \bibfield  {author} {\bibinfo {author} {\bibfnamefont {W.~K.}\ \bibnamefont
  {Tse}}\ and\ \bibinfo {author} {\bibfnamefont {S.}~\bibnamefont
  {Das~Sarma}},\ }\href@noop {} {\bibfield  {journal} {\bibinfo  {journal}
  {Phys. Rev. Lett.}\ }\textbf {\bibinfo {volume} {96}},\ \bibinfo {pages}
  {56601} (\bibinfo {year} {2006})}\BibitemShut {NoStop}%
\bibitem [{\citenamefont {Ploog}(1987)}]{Ploog1987}%
  \BibitemOpen
  \bibfield  {author} {\bibinfo {author} {\bibfnamefont {K.}~\bibnamefont
  {Ploog}},\ }\href@noop {} {\bibfield  {journal} {\bibinfo  {journal} {J.
  Cryst. Growth}\ }\textbf {\bibinfo {volume} {81}},\ \bibinfo {pages} {304}
  (\bibinfo {year} {1987})}\BibitemShut {NoStop}%
\bibitem [{\citenamefont {Haug}\ \emph {et~al.}(1987)\citenamefont {Haug},
  \citenamefont {Gerhardts}, \citenamefont {von Klitzing},\ and\ \citenamefont
  {Ploog}}]{Haug1987}%
  \BibitemOpen
  \bibfield  {author} {\bibinfo {author} {\bibfnamefont {R.~J.}\ \bibnamefont
  {Haug}}, \bibinfo {author} {\bibfnamefont {R.~R.}\ \bibnamefont {Gerhardts}},
  \bibinfo {author} {\bibfnamefont {K.}~\bibnamefont {von Klitzing}}, \ and\
  \bibinfo {author} {\bibfnamefont {K.}~\bibnamefont {Ploog}},\ }\href@noop {}
  {\bibfield  {journal} {\bibinfo  {journal} {Phys. Rev. Lett.}\ }\textbf
  {\bibinfo {volume} {59}},\ \bibinfo {pages} {1349} (\bibinfo {year}
  {1987})}\BibitemShut {NoStop}%
\bibitem [{\citenamefont {Hankiewicz}\ \emph {et~al.}(2006)\citenamefont
  {Hankiewicz}, \citenamefont {Vignale},\ and\ \citenamefont
  {Flatt{\'e}}}]{Hankiewicz2006PRL}%
  \BibitemOpen
  \bibfield  {author} {\bibinfo {author} {\bibfnamefont {E.~M.}\ \bibnamefont
  {Hankiewicz}}, \bibinfo {author} {\bibfnamefont {G.}~\bibnamefont {Vignale}},
  \ and\ \bibinfo {author} {\bibfnamefont {M.~E.}\ \bibnamefont {Flatt{\'e}}},\
  }\href@noop {} {\bibfield  {journal} {\bibinfo  {journal} {Phys. Rev. Lett.}\
  }\textbf {\bibinfo {volume} {97}},\ \bibinfo {pages} {266601} (\bibinfo
  {year} {2006})}\BibitemShut {NoStop}%
\bibitem [{\citenamefont {Mott}(1929)}]{Mott1929}%
  \BibitemOpen
  \bibfield  {author} {\bibinfo {author} {\bibfnamefont {N.~F.}\ \bibnamefont
  {Mott}},\ }\href@noop {} {\bibfield  {journal} {\bibinfo  {journal} {Proc. R.
  Soc. A}\ }\textbf {\bibinfo {volume} {124}},\ \bibinfo {pages} {425}
  (\bibinfo {year} {1929})}\BibitemShut {NoStop}%
\bibitem [{\citenamefont {Smit}(1955)}]{Smit1955}%
  \BibitemOpen
  \bibfield  {author} {\bibinfo {author} {\bibfnamefont {J.}~\bibnamefont
  {Smit}},\ }\href@noop {} {\bibfield  {journal} {\bibinfo  {journal}
  {Physica}\ }\textbf {\bibinfo {volume} {21}},\ \bibinfo {pages} {877}
  (\bibinfo {year} {1955})}\BibitemShut {NoStop}%
\bibitem [{\citenamefont {Smit}(1958)}]{Smit1958}%
  \BibitemOpen
  \bibfield  {author} {\bibinfo {author} {\bibfnamefont {J.}~\bibnamefont
  {Smit}},\ }\href@noop {} {\bibfield  {journal} {\bibinfo  {journal}
  {Physica}\ }\textbf {\bibinfo {volume} {24}},\ \bibinfo {pages} {39}
  (\bibinfo {year} {1958})}\BibitemShut {NoStop}%
\bibitem [{\citenamefont {Berger}(1970)}]{Berger1970}%
  \BibitemOpen
  \bibfield  {author} {\bibinfo {author} {\bibfnamefont {L.}~\bibnamefont
  {Berger}},\ }\href@noop {} {\bibfield  {journal} {\bibinfo  {journal} {Phys.
  Rev. B}\ }\textbf {\bibinfo {volume} {2}},\ \bibinfo {pages} {4559} (\bibinfo
  {year} {1970})}\BibitemShut {NoStop}%
\bibitem [{\citenamefont {Berger}(1972)}]{Berger1972}%
  \BibitemOpen
  \bibfield  {author} {\bibinfo {author} {\bibfnamefont {L.}~\bibnamefont
  {Berger}},\ }\href@noop {} {\bibfield  {journal} {\bibinfo  {journal} {Phys.
  Rev. B}\ }\textbf {\bibinfo {volume} {5}},\ \bibinfo {pages} {1862} (\bibinfo
  {year} {1972})}\BibitemShut {NoStop}%
\bibitem [{\citenamefont {Lyo}\ and\ \citenamefont
  {Holstein}(1972)}]{Lyo-Holstein1972}%
  \BibitemOpen
  \bibfield  {author} {\bibinfo {author} {\bibfnamefont {S.~K.}\ \bibnamefont
  {Lyo}}\ and\ \bibinfo {author} {\bibfnamefont {T.}~\bibnamefont {Holstein}},\
  }\href@noop {} {\bibfield  {journal} {\bibinfo  {journal} {Phys. Rev. Lett.}\
  }\textbf {\bibinfo {volume} {29}},\ \bibinfo {pages} {423} (\bibinfo {year}
  {1972})}\BibitemShut {NoStop}%
\bibitem [{\citenamefont {Karplus}\ and\ \citenamefont
  {Luttinger}(1954)}]{Karplus-Luttinger1954}%
  \BibitemOpen
  \bibfield  {author} {\bibinfo {author} {\bibfnamefont {R.}~\bibnamefont
  {Karplus}}\ and\ \bibinfo {author} {\bibfnamefont {J.~M.}\ \bibnamefont
  {Luttinger}},\ }\href@noop {} {\bibfield  {journal} {\bibinfo  {journal}
  {Phys. Rev.}\ }\textbf {\bibinfo {volume} {95}},\ \bibinfo {pages} {1154}
  (\bibinfo {year} {1954})}\BibitemShut {NoStop}%
\bibitem [{\citenamefont {Luttinger}(1958)}]{Luttinger1958}%
  \BibitemOpen
  \bibfield  {author} {\bibinfo {author} {\bibfnamefont {J.~M.}\ \bibnamefont
  {Luttinger}},\ }\href@noop {} {\bibfield  {journal} {\bibinfo  {journal}
  {Phys. Rev.}\ }\textbf {\bibinfo {volume} {112}},\ \bibinfo {pages} {739}
  (\bibinfo {year} {1958})}\BibitemShut {NoStop}%
\bibitem [{\citenamefont {Nozi\`eres}\ and\ \citenamefont
  {Lewiner}(1973)}]{Nozieres1973}%
  \BibitemOpen
  \bibfield  {author} {\bibinfo {author} {\bibfnamefont {P.}~\bibnamefont
  {Nozi\`eres}}\ and\ \bibinfo {author} {\bibfnamefont {C.}~\bibnamefont
  {Lewiner}},\ }\href {\doibase 10.1051/jphys:019730034010090100} {\bibfield
  {journal} {\bibinfo  {journal} {J. Phys. (Paris)}\ }\textbf {\bibinfo
  {volume} {34}},\ \bibinfo {pages} {901} (\bibinfo {year} {1973})}\BibitemShut
  {NoStop}%
\bibitem [{\citenamefont {Nagaosa}\ \emph {et~al.}(2010)\citenamefont
  {Nagaosa}, \citenamefont {Sinova}, \citenamefont {Onoda}, \citenamefont
  {MacDonald},\ and\ \citenamefont {Ong}}]{Nagaosa2010}%
  \BibitemOpen
  \bibfield  {author} {\bibinfo {author} {\bibfnamefont {N.}~\bibnamefont
  {Nagaosa}}, \bibinfo {author} {\bibfnamefont {J.}~\bibnamefont {Sinova}},
  \bibinfo {author} {\bibfnamefont {S.}~\bibnamefont {Onoda}}, \bibinfo
  {author} {\bibfnamefont {A.~H.}\ \bibnamefont {MacDonald}}, \ and\ \bibinfo
  {author} {\bibfnamefont {N.~P.}\ \bibnamefont {Ong}},\ }\href@noop {}
  {\bibfield  {journal} {\bibinfo  {journal} {Rev. Mod. Phys.}\ }\textbf
  {\bibinfo {volume} {82}},\ \bibinfo {pages} {1539} (\bibinfo {year}
  {2010})}\BibitemShut {NoStop}%
\bibitem [{\citenamefont {Johnson}\ and\ \citenamefont
  {Silsbee}(1987)}]{Johnson1987}%
  \BibitemOpen
  \bibfield  {author} {\bibinfo {author} {\bibfnamefont {M.}~\bibnamefont
  {Johnson}}\ and\ \bibinfo {author} {\bibfnamefont {R.~H.}\ \bibnamefont
  {Silsbee}},\ }\href@noop {} {\bibfield  {journal} {\bibinfo  {journal} {Phys.
  Rev. B}\ }\textbf {\bibinfo {volume} {35}},\ \bibinfo {pages} {4959}
  (\bibinfo {year} {1987})}\BibitemShut {NoStop}%
\bibitem [{\citenamefont {Bauer}\ \emph {et~al.}(2012)\citenamefont {Bauer},
  \citenamefont {Saitoh},\ and\ \citenamefont {van Wees}}]{Bauer2012}%
  \BibitemOpen
  \bibfield  {author} {\bibinfo {author} {\bibfnamefont {G.~E.~W.}\
  \bibnamefont {Bauer}}, \bibinfo {author} {\bibfnamefont {E.}~\bibnamefont
  {Saitoh}}, \ and\ \bibinfo {author} {\bibfnamefont {B.~J.}\ \bibnamefont {van
  Wees}},\ }\href@noop {} {\bibfield  {journal} {\bibinfo  {journal} {Nat.
  Mater.}\ }\textbf {\bibinfo {volume} {11}},\ \bibinfo {pages} {391} (\bibinfo
  {year} {2012})}\BibitemShut {NoStop}%
\bibitem [{\citenamefont {Smith}(1911)}]{Smith1911}%
  \BibitemOpen
  \bibfield  {author} {\bibinfo {author} {\bibfnamefont {A.~W.}\ \bibnamefont
  {Smith}},\ }\href@noop {} {\bibfield  {journal} {\bibinfo  {journal} {Phys.
  Rev. (Series I)}\ }\textbf {\bibinfo {volume} {33}},\ \bibinfo {pages} {295}
  (\bibinfo {year} {1911})}\BibitemShut {NoStop}%
\bibitem [{\citenamefont {Kondorskii}\ and\ \citenamefont
  {Vasileva}(1964)}]{Kondorskii_Vasileva1964}%
  \BibitemOpen
  \bibfield  {author} {\bibinfo {author} {\bibfnamefont {E.~I.}\ \bibnamefont
  {Kondorskii}}\ and\ \bibinfo {author} {\bibfnamefont {R.~P.}\ \bibnamefont
  {Vasileva}},\ }\href@noop {} {\bibfield  {journal} {\bibinfo  {journal} {Sov.
  Phys. JEPT}\ }\textbf {\bibinfo {volume} {18}},\ \bibinfo {pages} {277}
  (\bibinfo {year} {1964})}\BibitemShut {NoStop}%
\bibitem [{\citenamefont {Kondorskii}(1964)}]{Kondorskii1964}%
  \BibitemOpen
  \bibfield  {author} {\bibinfo {author} {\bibfnamefont {E.~I.}\ \bibnamefont
  {Kondorskii}},\ }\href@noop {} {\bibfield  {journal} {\bibinfo  {journal}
  {Sov. Phys. JETP}\ }\textbf {\bibinfo {volume} {18}},\ \bibinfo {pages} {351}
  (\bibinfo {year} {1964})}\BibitemShut {NoStop}%
\bibitem [{\citenamefont {Seki}\ \emph {et~al.}(2010)\citenamefont {Seki},
  \citenamefont {Sugai}, \citenamefont {Hasegawa}, \citenamefont {Mitani},\
  and\ \citenamefont {Takanashi}}]{Seki2010}%
  \BibitemOpen
  \bibfield  {author} {\bibinfo {author} {\bibfnamefont {T.}~\bibnamefont
  {Seki}}, \bibinfo {author} {\bibfnamefont {I.}~\bibnamefont {Sugai}},
  \bibinfo {author} {\bibfnamefont {Y.}~\bibnamefont {Hasegawa}}, \bibinfo
  {author} {\bibfnamefont {S.}~\bibnamefont {Mitani}}, \ and\ \bibinfo {author}
  {\bibfnamefont {K.}~\bibnamefont {Takanashi}},\ }\href@noop {} {\bibfield
  {journal} {\bibinfo  {journal} {Solid State Commun.}\ }\textbf {\bibinfo
  {volume} {150}},\ \bibinfo {pages} {496} (\bibinfo {year}
  {2010})}\BibitemShut {NoStop}%
\bibitem [{\citenamefont {Cheng}\ \emph {et~al.}(2008)\citenamefont {Cheng},
  \citenamefont {Xing}, \citenamefont {Sun},\ and\ \citenamefont
  {Xie}}]{Cheng2008}%
  \BibitemOpen
  \bibfield  {author} {\bibinfo {author} {\bibfnamefont {S.}~\bibnamefont
  {Cheng}}, \bibinfo {author} {\bibfnamefont {Y.}~\bibnamefont {Xing}},
  \bibinfo {author} {\bibfnamefont {Q.}~\bibnamefont {Sun}}, \ and\ \bibinfo
  {author} {\bibfnamefont {X.~C.}\ \bibnamefont {Xie}},\ }\href@noop {}
  {\bibfield  {journal} {\bibinfo  {journal} {Phys. Rev. B}\ }\textbf {\bibinfo
  {volume} {78}},\ \bibinfo {pages} {045302} (\bibinfo {year}
  {2008})}\BibitemShut {NoStop}%
\bibitem [{\citenamefont {Ma}(2010)}]{Ma2010}%
  \BibitemOpen
  \bibfield  {author} {\bibinfo {author} {\bibfnamefont {Z.}~\bibnamefont
  {Ma}},\ }\href@noop {} {\bibfield  {journal} {\bibinfo  {journal} {Solid
  State Commun.}\ }\textbf {\bibinfo {volume} {150}},\ \bibinfo {pages} {510}
  (\bibinfo {year} {2010})}\BibitemShut {NoStop}%
\bibitem [{\citenamefont {Liu}\ and\ \citenamefont {Xie}(2010)}]{Liu2010}%
  \BibitemOpen
  \bibfield  {author} {\bibinfo {author} {\bibfnamefont {X.}~\bibnamefont
  {Liu}}\ and\ \bibinfo {author} {\bibfnamefont {X.~C.}\ \bibnamefont {Xie}},\
  }\href@noop {} {\bibfield  {journal} {\bibinfo  {journal} {Solid State
  Commun.}\ }\textbf {\bibinfo {volume} {150}},\ \bibinfo {pages} {471}
  (\bibinfo {year} {2010})}\BibitemShut {NoStop}%
\bibitem [{\citenamefont {Elliott}(1954)}]{Elliott1954}%
  \BibitemOpen
  \bibfield  {author} {\bibinfo {author} {\bibfnamefont {R.~J.}\ \bibnamefont
  {Elliott}},\ }\href@noop {} {\bibfield  {journal} {\bibinfo  {journal} {Phys.
  Rev.}\ }\textbf {\bibinfo {volume} {96}},\ \bibinfo {pages} {266} (\bibinfo
  {year} {1954})}\BibitemShut {NoStop}%
\bibitem [{\citenamefont {Yafet}(1963)}]{Yafet1963}%
  \BibitemOpen
  \bibfield  {author} {\bibinfo {author} {\bibfnamefont {Y.}~\bibnamefont
  {Yafet}},\ }in\ \href {\doibase 10.1016/S0081-1947(08)60259-3} {\emph
  {\bibinfo {booktitle} {Solid State Physics}}},\ Vol.~\bibinfo {volume} {14},\
  \bibinfo {editor} {edited by\ \bibinfo {editor} {\bibfnamefont
  {F.}~\bibnamefont {Seitz}}\ and\ \bibinfo {editor} {\bibfnamefont
  {D.}~\bibnamefont {Turnbull}}}\ (\bibinfo  {publisher} {Academic, New York},\
  \bibinfo {year} {1963})\ pp.\ \bibinfo {pages} {1 -- 98}\BibitemShut
  {NoStop}%
\bibitem [{\citenamefont {Averkiev}\ \emph {et~al.}(2002)\citenamefont
  {Averkiev}, \citenamefont {Golub},\ and\ \citenamefont
  {Willander}}]{Averkiev2002}%
  \BibitemOpen
  \bibfield  {author} {\bibinfo {author} {\bibfnamefont {N.~S.}\ \bibnamefont
  {Averkiev}}, \bibinfo {author} {\bibfnamefont {L.~E.}\ \bibnamefont {Golub}},
  \ and\ \bibinfo {author} {\bibfnamefont {M.}~\bibnamefont {Willander}},\
  }\href@noop {} {\bibfield  {journal} {\bibinfo  {journal} {J. Phys.: Condens.
  Matter}\ }\textbf {\bibinfo {volume} {14}},\ \bibinfo {pages} {R271}
  (\bibinfo {year} {2002})}\BibitemShut {NoStop}%
\bibitem [{\citenamefont {Bronold}\ \emph {et~al.}(2004)\citenamefont
  {Bronold}, \citenamefont {Saxena},\ and\ \citenamefont
  {Smith}}]{Bronold2004}%
  \BibitemOpen
  \bibfield  {author} {\bibinfo {author} {\bibfnamefont {F.~X.}\ \bibnamefont
  {Bronold}}, \bibinfo {author} {\bibfnamefont {A.}~\bibnamefont {Saxena}}, \
  and\ \bibinfo {author} {\bibfnamefont {D.~L.}\ \bibnamefont {Smith}},\
  }\href@noop {} {\bibfield  {journal} {\bibinfo  {journal} {Phys. Rev. B}\
  }\textbf {\bibinfo {volume} {70}},\ \bibinfo {pages} {245210} (\bibinfo
  {year} {2004})}\BibitemShut {NoStop}%
\bibitem [{\citenamefont {Landau}\ and\ \citenamefont
  {Lifshitz}(1965)}]{Landau1965}%
  \BibitemOpen
  \bibfield  {author} {\bibinfo {author} {\bibfnamefont {L.~D.}\ \bibnamefont
  {Landau}}\ and\ \bibinfo {author} {\bibfnamefont {E.~M.}\ \bibnamefont
  {Lifshitz}},\ }\href@noop {} {\emph {\bibinfo {title} {Quantum mechanics,
  Course of theoretical physics}}}\ (\bibinfo  {publisher} {Pergamon Press, New
  York},\ \bibinfo {year} {1965})\BibitemShut {NoStop}%
\bibitem [{\citenamefont {Culcer}\ \emph {et~al.}(2010)\citenamefont {Culcer},
  \citenamefont {Hankiewicz}, \citenamefont {Vignale},\ and\ \citenamefont
  {Winkler}}]{Culcer2010}%
  \BibitemOpen
  \bibfield  {author} {\bibinfo {author} {\bibfnamefont {D.}~\bibnamefont
  {Culcer}}, \bibinfo {author} {\bibfnamefont {E.~M.}\ \bibnamefont
  {Hankiewicz}}, \bibinfo {author} {\bibfnamefont {G.}~\bibnamefont {Vignale}},
  \ and\ \bibinfo {author} {\bibfnamefont {R.}~\bibnamefont {Winkler}},\
  }\href@noop {} {\bibfield  {journal} {\bibinfo  {journal} {Phys. Rev. B}\
  }\textbf {\bibinfo {volume} {81}},\ \bibinfo {pages} {125332} (\bibinfo
  {year} {2010})}\BibitemShut {NoStop}%
\bibitem [{\citenamefont {Callen}(1960)}]{Callen1960thermodynamics}%
  \BibitemOpen
  \bibfield  {author} {\bibinfo {author} {\bibfnamefont {H.~B.}\ \bibnamefont
  {Callen}},\ }\href {http://books.google.co.jp/books?id=v10wAAAAIAAJ} {\emph
  {\bibinfo {title} {Thermodynamics}}}\ (\bibinfo  {publisher} {Wiley, New
  York},\ \bibinfo {year} {1960})\BibitemShut {NoStop}%
\bibitem [{\citenamefont {Groot}\ and\ \citenamefont
  {Mazur}(1962)}]{Groot1962nonequilibrium}%
  \BibitemOpen
  \bibfield  {author} {\bibinfo {author} {\bibfnamefont {S.~R.}\ \bibnamefont
  {Groot}}\ and\ \bibinfo {author} {\bibfnamefont {P.}~\bibnamefont {Mazur}},\
  }\href {http://books.google.co.jp/books?id=3b-wAAAAIAAJ} {\emph {\bibinfo
  {title} {Non-equilibrium thermodynamics}}}\ (\bibinfo  {publisher}
  {North-Holland, Amsterdam},\ \bibinfo {year} {1962})\BibitemShut {NoStop}%
\bibitem [{\citenamefont {Mott}\ and\ \citenamefont
  {Jones}(1936)}]{Mott-Jones}%
  \BibitemOpen
  \bibfield  {author} {\bibinfo {author} {\bibfnamefont {N.~F.}\ \bibnamefont
  {Mott}}\ and\ \bibinfo {author} {\bibfnamefont {H.}~\bibnamefont {Jones}},\
  }\href {http://books.google.co.jp/books?id=-qtBAAAAIAAJ} {\emph {\bibinfo
  {title} {The theory of the properties of metals and alloys}}}\ (\bibinfo
  {publisher} {Clarendon, Oxford},\ \bibinfo {year} {1936})\BibitemShut
  {NoStop}%
\bibitem [{\citenamefont {Onsager}(1931{\natexlab{a}})}]{Onsager1931_I}%
  \BibitemOpen
  \bibfield  {author} {\bibinfo {author} {\bibfnamefont {L.}~\bibnamefont
  {Onsager}},\ }\href@noop {} {\bibfield  {journal} {\bibinfo  {journal} {Phys.
  Rev.}\ }\textbf {\bibinfo {volume} {37}},\ \bibinfo {pages} {405} (\bibinfo
  {year} {1931}{\natexlab{a}})}\BibitemShut {NoStop}%
\bibitem [{\citenamefont {Onsager}(1931{\natexlab{b}})}]{Onsager1931_II}%
  \BibitemOpen
  \bibfield  {author} {\bibinfo {author} {\bibfnamefont {L.}~\bibnamefont
  {Onsager}},\ }\href@noop {} {\bibfield  {journal} {\bibinfo  {journal} {Phys.
  Rev.}\ }\textbf {\bibinfo {volume} {38}},\ \bibinfo {pages} {2265} (\bibinfo
  {year} {1931}{\natexlab{b}})}\BibitemShut {NoStop}%
\bibitem [{\citenamefont {Hankiewicz}\ and\ \citenamefont
  {Vignale}(2009)}]{Hankiewicz2009}%
  \BibitemOpen
  \bibfield  {author} {\bibinfo {author} {\bibfnamefont {E.~M.}\ \bibnamefont
  {Hankiewicz}}\ and\ \bibinfo {author} {\bibfnamefont {G.}~\bibnamefont
  {Vignale}},\ }\href {http://stacks.iop.org/0953-8984/21/i=25/a=253202}
  {\bibfield  {journal} {\bibinfo  {journal} {J. Phys.: Condens. Matter}\
  }\textbf {\bibinfo {volume} {21}},\ \bibinfo {pages} {253202} (\bibinfo
  {year} {2009})}\BibitemShut {NoStop}%
\bibitem [{\citenamefont {Kohn}\ and\ \citenamefont
  {Luttinger}(1957)}]{Kohn-Luttinger1957}%
  \BibitemOpen
  \bibfield  {author} {\bibinfo {author} {\bibfnamefont {W.}~\bibnamefont
  {Kohn}}\ and\ \bibinfo {author} {\bibfnamefont {J.~M.}\ \bibnamefont
  {Luttinger}},\ }\href@noop {} {\bibfield  {journal} {\bibinfo  {journal}
  {Phys. Rev.}\ }\textbf {\bibinfo {volume} {108}},\ \bibinfo {pages} {590}
  (\bibinfo {year} {1957})}\BibitemShut {NoStop}%
\end{thebibliography}
\end{document}